\documentclass[preprint, superscriptaddress, amsmath,amssymb,aps,physrev]{revtex4-1}

\usepackage{cancel}

\usepackage[dvipsnames]{xcolor}
\usepackage{graphicx}
\usepackage{dcolumn}
\usepackage{bm}
\usepackage{ulem}
\usepackage{hyperref}
\hypersetup{colorlinks=true,urlcolor=blue,citecolor=blue}
\usepackage[symbol]{footmisc}

\definecolor{fg}{rgb}{1, 0, 0}

\definecolor{fb}{rgb}{0, 0, 1}

\definecolor{remove}{rgb}{0, 1, 0}

\begin{document}

\title{Control of intervalley scattering in Bi$_2$Te$_3$ \\ via temperature-dependent band renormalization }

\author{A. Jabed}
\altaffiliation{These authors equally contributed}
\affiliation{Advanced Laser Light Source, Institut National de la Recherche Scientifique, Varennes QC J3X 1S2 Canada}
\affiliation{Université de Bordeaux-CNRS-CEA,CELIA,UMR5107,F33405 Talence,France}

\author{F. Goto}
\altaffiliation{These authors equally contributed}
\affiliation{Advanced Laser Light Source, Institut National de la Recherche Scientifique, Varennes QC J3X 1S2 Canada}
\author{B. Frimpong}
\author{D. Armanno}
\author{A. Longa}
\affiliation{Advanced Laser Light Source, Institut National de la Recherche Scientifique, Varennes QC J3X 1S2 Canada}
\author{M. Michiardi}
\author{A. Damascelli}
\affiliation{Quantum Matter Institute, University of British Columbia, Vancouver, BC V6T 1Z4, Canada}
\affiliation{Department of Physics $\&$ Astronomy, University of British Columbia, Vancouver, BC V6T 1Z1, Canada}
\author{P. Hofmann}
\affiliation{Department of Physics and Astronomy, Interdisciplinary Nanoscience Center, Aarhus University, 8000 Aarhus C, Denmark}
\author{G. Jargot}
\author{H. Ibrahim}
\author{F. L\'{e}gar\'{e}}
\author{N. Gauthier}
\affiliation{Advanced Laser Light Source, Institut National de la Recherche Scientifique, Varennes QC J3X 1S2 Canada}
\author{S. Beaulieu}
\affiliation{Université de Bordeaux-CNRS-CEA,CELIA,UMR5107,F33405 Talence,France}
\author{F. Boschini}
\email[]{fabio.boschini@inrs.ca}
\affiliation{Advanced Laser Light Source, Institut National de la Recherche Scientifique, Varennes QC J3X 1S2 Canada}
\affiliation{Quantum Matter Institute, University of British Columbia, Vancouver, BC V6T 1Z4, Canada}

\date{\today}

\begin{abstract}
The control of out-of-equilibrium electron dynamics in topological insulators is essential to unlock their potential in next-generation quantum technologies. However, the role of temperature on the renormalization of the electronic band structure and, consequently, on electron scattering processes is still elusive. Here, using high-resolution time- and angle-resolved photoemission spectroscopy (TR-ARPES), we show that even a modest ($\sim$15\,meV) renormalization of the conduction band of Bi$_2$Te$_3$ can critically affect bulk and surface electron scattering processes. Supported by a kinetic Monte Carlo toy-model, we show that temperature-induced changes in the bulk band structure modulate the intervalley electron-phonon scattering rate, reshaping the out-of-equilibrium response. This work establishes temperature as an effective control knob for engineering scattering pathways in topological insulators.
\end{abstract}
\maketitle
\section*{Introduction}
Three-dimensional topological insulators (TIs), owing to the presence of a topologically-protected metallic surface state (TSS) within an insulating bulk energy gap \cite{Fu_2007,Moore_2010,Chen_2009,Hasan_2010,Ferreira_2013,Jin_2023}, are still prime candidates for advanced technological applications, such as photogalvanic current generation \cite{tao2020pure}, (opto-)spintronics \cite{Haldane_2017,Hsieh2009ObservationInsulators,michiardi2022optical}, and quantum computing \cite{Qi_2011,Legg_2021,Paudel_2013}. In this regard, the common denominator of all TI-based devices is the demand for precise control of their electronic properties and response to external stimuli, in both perturbative and non-perturbative regimes.  
This desire is contingent upon the need for a deeper understanding of their out-of-equilibrium charge dynamics. 

To this end, time- and angle-resolved photoemission spectroscopy (TR-ARPES) is nowadays a well-established tool for probing ultrafast electron dynamics in quantum materials with remarkable temporal, energy and momentum resolutions \cite{Boschini2024,Zonno2021}. 
When applied to TIs, TR-ARPES has provided direct evidence for electron relaxation processes in the TSS well beyond the timescale of a few ps \cite{Crepaldi_2013,Sobota2012,Sterzi_2017,Neupane2015,Huang2023}, results that have been discussed in terms of the intrinsically weak electron-phonon coupling caused by the limited scattering phase space in concert with the topological protection of the TSS against large-momentum scattering
\cite{Barriga_2016,Michiardi2014,Hajlaoui2014,Sobota2014Distinguishing}. 
However, although it is widely accepted that bulk bands act as charge reservoirs for the TSS \cite{Barriga2017,Hajlaoui2012,Papalazarou_2018,Soifer_2019}, it is still unclear how their specific electronic dispersion may impact ultrafast electron relaxation dynamics. In this context, Chen $et\ al.$ \cite{chen2024distinct} have shown how ultrafast scattering processes in p-doped Bi$_2$Te$_3$ depend closely on the pump photon energy, \textit{i.e.} on the specific unoccupied states that are populated by photoexcited electrons. In particular, they reported that photoexcitation with 330\,meV photon energy results in slow carrier dynamics dominated by intervalley scattering.

Here, we present a high-resolution TR-ARPES study of p-doped Bi$_2$Te$_3$ and, with the support of kinetic Monte Carlo (KMC) simulations, we offer first evidence of how a minor renormalization of the bulk bands dispersion has a substantial impact on the ultrafast electron dynamics. Indeed, as the sample temperature increases, the interplay between thermal expansion and electron-phonon coupling results in a renormalization of the conduction band (CB), opening a new pathway for efficient intervalley scattering towards the center of the Brillouin zone ($\Gamma$) \cite{Monserrat2016}. Our experimental evidence is reminiscent of 
what is found in semiconducting transition metal dichalcogenides, where the charge dynamics is dominated by intra- and inter-valley scattering processes \cite{Tanimura_2021,Sjakste_2018,Tanimura_2016,Kanasaki_2014, Tanimura_2015}, and the relative energy position of the different valleys influences population lifetime \cite{ataei2021competitive,ulstrup2016ultrafast,lee2021direct}.
Our results underscore
the crucial, and fairly unexplored, combined role of temperature and the bulk band structure in feeding carriers to the technologically relevant TSS. Indeed, changes in the CB dispersion come hand-in-hand with a reduction of the direct bulk gap, \textit{i.e.} at the $\Gamma$ point, as the temperature increases, which may be key information for capturing the thermoelectric properties of Bi$_2$Te$_3$ \cite{Rittweger2014,Liang2016,Cao2023,Monserrat2016}.

\section*{Results}
\subsection*{Time-resolved ARPES mapping of the unoccupied states of p-doped Bi$_2$Te$_3$}
\begin{figure*}[hb]
    \centering
    \includegraphics[width=0.99\linewidth]{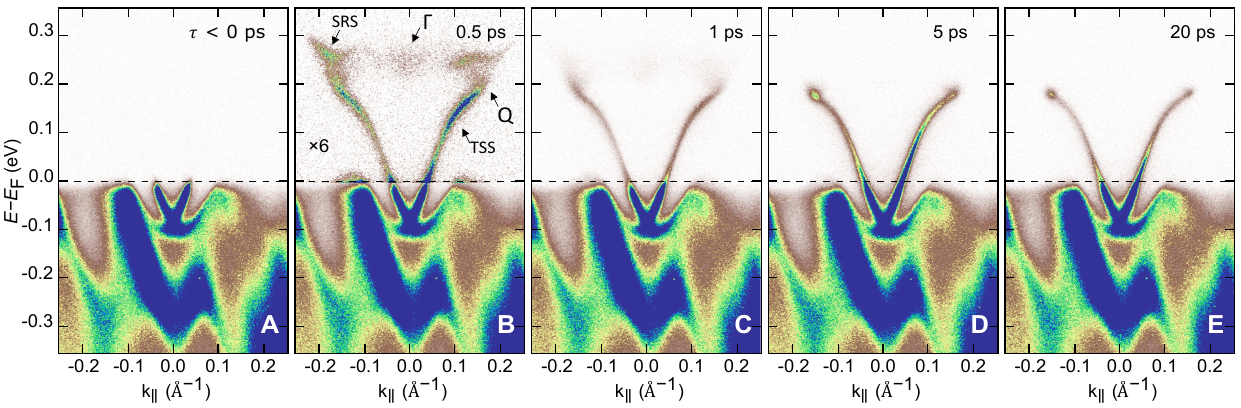}
   \caption{\textbf{Long-lasting intensity buildup in the unoccupied states of p-doped Bi$_2$Te$_3$}. Spectra acquired (6\,eV probe\,/\,300\,meV pump) at a temperature of 30\,K, along the $\Gamma$--M direction, for different pump-probe delays: (A) $\tau<0$ ps, (B) 0.5 ps, (C) 1 ps, (D) 5 ps and (E) 20 ps. See Supplementary Text for details of the experimental geometry.}
    \label{fig1}
\end{figure*}

Figure\,\ref{fig1} displays TR-ARPES maps acquired at 30 K using 6 eV probe and 300\,meV pump pulses (details in Methods), along the $\Gamma$-M direction, for different pump-probe delays (additional data at different temperatures are provided in Fig.\,\ref{fig:Fig_SI2}). In agreement with Ref.\,\onlinecite{chen2024distinct}, the 300\,meV pump excitation enables direct promotion of electrons from the valence band (VB) into the CB of Bi$_2$Te$_3$.  In particular, we report a direct optical transition into the surface resonance state (SRS) \cite{Hedayat2021,cacho2015momentum,jozwiak2016spin}, slightly off the $\Gamma$ point. 

By simple visual inspection of the ARPES map at 0.5\,ps pump-probe delay (Fig.\,\ref{fig1}B), it is clear that the pump pulse populates a continuum of states connecting the CB absolute minimum at $\pm0.15 \ \text{\AA}^{-1}$(Q-valleys) and the CB local minimum at the $\Gamma$ point ($\Gamma$-valley). While the CB signal at $\Gamma$ decays within a few ps, an intensity buildup forms and persists for tens of ps at the bottom of the Q-valleys, where the TSS merges with the CB. This intensity buildup has been previously reported and attributed to the first evidence of a spatially indirect topological exciton that forms by binding surface electrons (sitting on the TSS) with bulk holes (at the VB maximum) \cite{Mori2023}.

We verified that this intensity buildup is also observed with 6.2 eV-probe / 1.55 eV-pump photon energies, as well as that it fades with increasing temperature (see Supplementary Text, Fig.\,\ref{fig:Fig_SI2} and Fig.\,\ref{fig:Fig_SI4}), as would also be expected in the exciton formation scenario \cite{Mori2023}.
However, as detailed below, we show that the observed temperature dependence is also consistent with the interpretation of these spectral features as bulk band states, and it can be directly linked to the efficiency of the intervalley scattering between bulk state valleys. 
We also note that when the TSS is in close proximity to the bulk states, it could possibly inherit their orbital character \cite{Mori2023}. In fact, (i) the intensity buildup displays a trigonal pattern reminiscent of the bulk states ($k_x$-$k_y$ iso-energy contour maps in Fig.\,\ref{fig:Fig_SI1}), and (ii) the linear dichroism maps (inspired by Ref.\,\onlinecite{Beaulieu_2021,Cao_2013,Min_2019}, see Fig.\,\ref{fig:Fig_SI1}) hint at different orbital characters between the intensity buildup and the surrounding TSS.
Based on the discussion above, and since our conclusions do not depend on the existence of the spatially indirect exciton, we adopt the simplest possible model. We then
refer to the intensity buildup as a direct signature of the carrier accumulation at the bottom of the Q-valley, and we establish its population density as a measure of the strength of the $\Gamma\leftrightarrow \text{Q}$ intervalley electron-phonon scattering channel. 

\subsection*{Temperature dependence of bulk band dispersion and ultrafast electron dynamics}
\begin{figure*}[ht]
    \centering
    \includegraphics[width=0.97\linewidth]{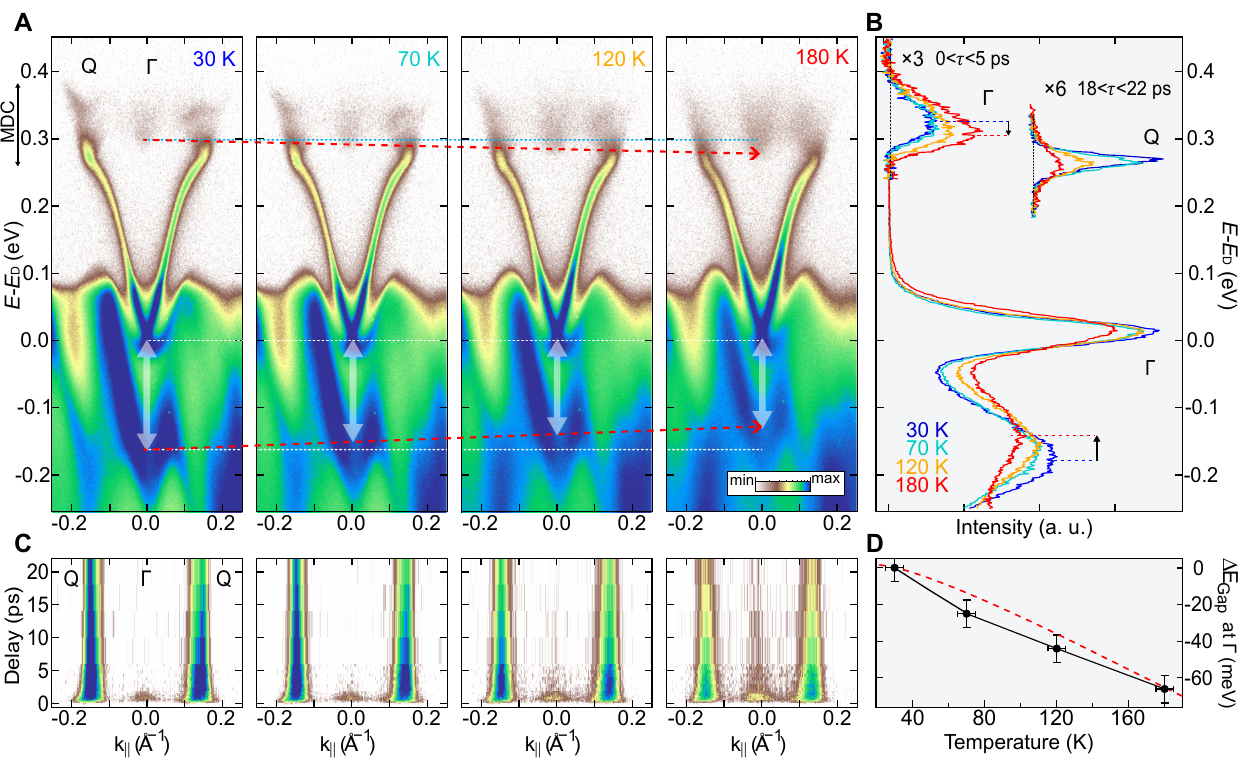}
    \caption{\textbf{Temperature-induced bulk band renormalization.} (A) Temperature-dependent TR-ARPES results on p-doped Bi$_2$Te$_3$, along the $\Gamma$--M direction, and integrated over the first 5\,ps. (B) Energy distribution curves at the $\Gamma$ point ($ k_{\parallel}\sim 0 \pm 50\, \text{m\AA}^{-1}$, up to 5 ps), and at the Q-valley ($ k_{\parallel}\sim -150 \pm 25\, \text{m\AA}^{-1} $, at 20\,ps). (C) Momentum distribution curves (MDCs) as a function of pump-probe delay at the corresponding temperatures. The integration window for MDCs is indicated by the double-headed black arrows in panel A. (D) Relative change of the direct bulk band gap at $\Gamma$ as a function of the temperature. The red dashed line displays ab-initio predictions, digitized from Ref.\,\cite{Monserrat2016}.}
    \label{fig2}
\end{figure*}

We now move on exploring the role of the temperature as tuning knob for the intervalley scattering in the CB.
To do so, Fig.\,\ref{fig2}A displays ARPES maps integrated over the first 5\,ps pump–probe delay at different temperatures (note that for comparison purposes, the energy axis is scaled with respect to the TSS Dirac point position, E$_{\text{D}}$). By focusing on the $\Gamma$ point, we report an upward (downward) shift in the energy of the VB (CB) maximum (minimum), thus highlighting a temperature-induced band structure renormalization. This shift is confirmed by inspecting the energy distribution curves (EDCs) at $\Gamma$ at different temperatures (Fig.\,\ref{fig2}B), and the temperature evolution of the bulk gap at $\Gamma$ (Fig.\,\ref{fig2}D). As the temperature increases from 30\,K to 180\,K, the band gap at $\Gamma$ reduces by $\sim$62\,meV, in good agreement with the \textit{ab-intio} theoretical estimate of Ref.\,\onlinecite{Monserrat2016} for bulk Bi$_2$Te$_3$, where both thermal expansion and electron-phonon coupling contributions have been accounted for (see red dashed line in Fig.\,\ref{fig2}D). In contrast, the peak position of the EDCs at the Q-valley, where the TSS is affected by trigonal-warping and merges into the CB \cite{Fu_2007}, remains essentially unchanged across all temperatures ($\sim$3 meV energy shift), as shown in Fig.\,\ref{fig2}B and Fig.\,\ref{fig3}A.

To capture how changes in the CB dispersion may impact the electron dynamics, Fig.\,\ref{fig2}C displays the transient evolution of the momentum distribution curves for four temperatures, in an energy range that includes the entire CB population (vertical line on the left of panel a). At low temperature, the $\Gamma$-valley spectral weight vanishes within a few ps and the Q-valley dominates the dynamics. As the temperature increases, a long-lasting slowly-decaying signal appears at the $\Gamma$-valley and the spectral weight at the Q-valleys is reduced, in good agreement with what is observed in the EDCs of Fig.\,\ref{fig2}B (see also Fig.\,\ref{fig3}D).

Figure\,\ref{fig2}C points towards a temperature-dependent coupling between the $\Gamma$- and Q-valleys. 
To be more quantitative, and identify a figure of merit that captures the role of the bulk band renormalization in defining electron dynamics, we compare normalized EDCs at the Q-valley (20 ps) and at the $\Gamma$-valley (0-5 ps delay range). In particular, we assess the spectral overlap between the $\Gamma$- and Q-valleys EDCs,illustrated by the shadow area in  Fig.\,\ref{fig3}A, by evaluating the product of the two normalized EDCs (EDC$ _{\Gamma \cap Q}$, colored circles in Fig.\,\ref{fig3}B). The increase in the spectral overlap corresponds to an enhancement of the scattering phase space available for an electron to escape a given valley by intervalley scattering. This is also supported by the fact that ratio of spectral intensities at $\Gamma$ and Q ($I^\Gamma / I^Q $, colored squares in Fig.\,\ref{fig3}B) follows the same trend in temperature as the spectral overlap EDC$_{\Gamma\cap Q}$, thus indicating a transfer of spectral weight from the Q- to the $\Gamma$-valley mediated by intervalley electron-phonon scattering.

\subsection*{Temperature-induced changes of the scattering phase space} 
\begin{figure*}[hb]
    \centering
    \includegraphics[width=0.98\linewidth]{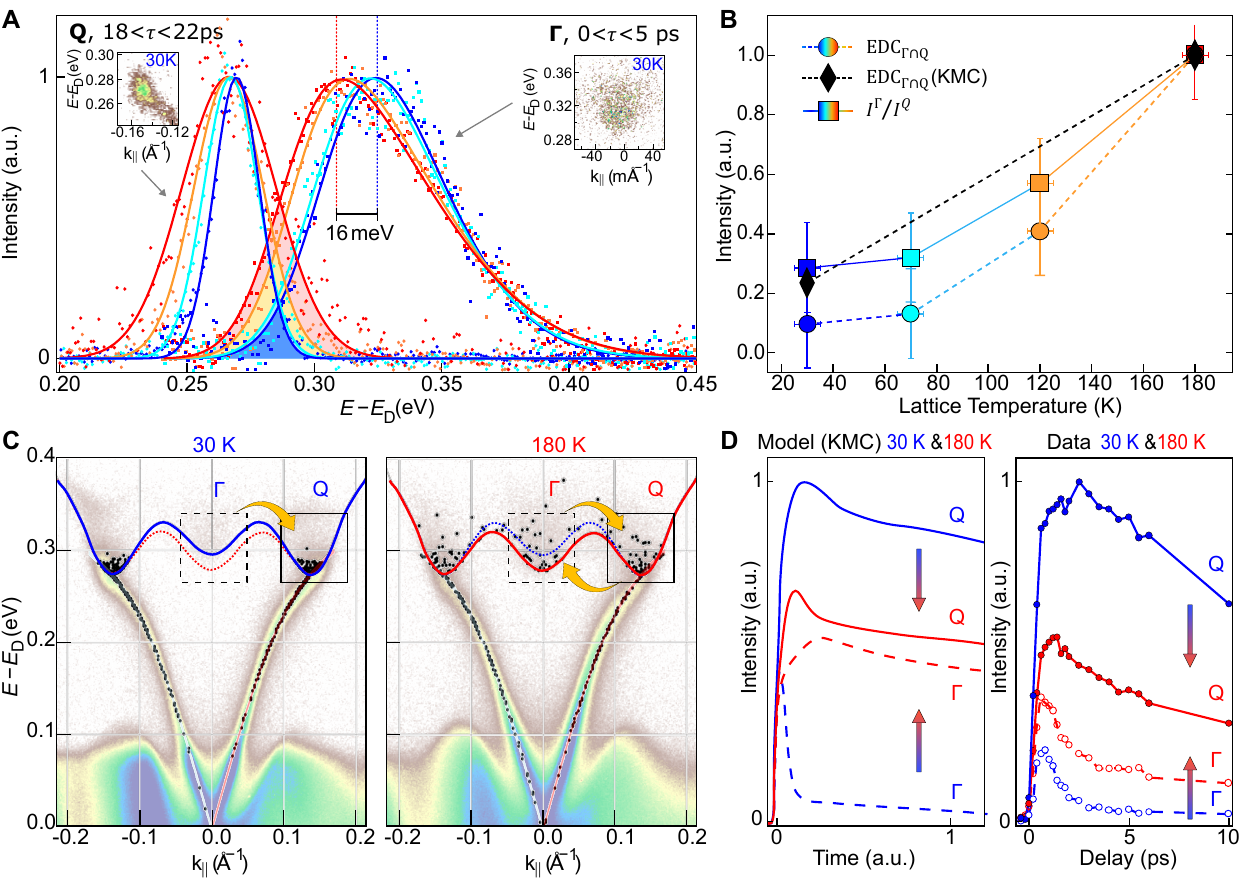}
    \caption{\textbf{Intervalley scattering behinds the emergence of the intensity buildup.} (A) Normalized Energy Distribution curves at different temperatures with superimposed an asymmetric gaussian fit for the $\Gamma$ (0$<\tau<$ 5ps) and Q valleys (18 ps$<\tau<$22 ps). The spectral overlap is indicated as shadow areas with different colors for each temperature. (B) The spectral overlap between the $\Gamma$- and Q-valley, EDC$_{\Gamma \cap Q}=I_{norm}^Q\times I_{norm}^\Gamma$ (colored circles), is compared with the ratio of the photoemission intensities for the two valleys (colored squares). (C) Electron occupancy obtained by the  electron-phonon scattering toy-model at 30 K and 180 K, for the last delay, superimposed on top of the experimental band structure from Fig.\,\ref{fig2}. (D) Comparison between simulated and experimental electron dynamics at $\Gamma$ (dashed lines) and Q (solid lines) valleys. The band structure renormalization increases the efficiency of intervalley scattering from Q to $\Gamma$.}
   \label{fig3}
   \end{figure*}

To better interpret our experimental evidence and the interplay between the $\Gamma$- and the Q-valleys, we implemented a toy-model for electron-phonon scattering using a kinetic Monte Carlo (KMC) approach, similar to Ref.\,\onlinecite{Sobota_2014}, but with momentum resolution \cite{na2020establishing}. We run KMC calculations for two different electronic band structures, namely at 30\,K and 180\,K (red and blue curves in Fig.\,\ref{fig3}C), that only differ by a small shift of $\sim$16\,meV of the CB minimum at $\Gamma$, based on the experimental observation (see Fig.\,\ref{fig2}A). We model the CB using a discrete distribution of equally spaced electronic states, and we tune the mesh density of the KMC model such that the spectral overlap between the KMC occupancies, $\Gamma \cap Q$, approximately follows the temperature dependence of the spectral overlap extracted by TR-ARPES data (see black diamonds in Fig.\,\ref{fig3}B). 

In the KMC calculations, we impose a Gaussian distribution of N electrons centered at 0.4\,eV above the Dirac point (to match the experimental condition at $\tau \sim$0.5\,ps, see Figs.\,\ref{fig1}-\ref{fig2}), and then evaluate the electron occupancy as a function of time, $f_i (t)$, by solving the rate equation:
\begin{equation}
  \frac{\partial f_i}{\partial t}=\sum_j [W_{j,i} f_j (1-f_i)-W_{i,j} f_i (1-f_j)],
  \label{eq:rate}
\end{equation}
where $W_{i,j}$ is the probability for each electron $(k_i,\,E_i)$ to have a transition into an unoccupied state $ (k_j,\,E_j)$ by absorption or emission of a phonon, provided by the Fermi's golden rule (see Methods).
Black circles in Fig.\,\ref{fig3}C display the electronic distributions obtained via KMC for 30\,K and 180\,K when a quasi-equilibrium condition is reached (see also Fig.\,\ref{fig:Fig_SI3} for different snapshots of the KMC toy-model). Fig.\,\ref{fig3}D offers also direct comparison between experimental and KMC time traces at $\Gamma$- and Q-valleys (integration boxes shown in Fig.\,\ref{fig3}C). 
Overall, this toy-model successfully captures (i) the formation of the characteristic long-standing intensity buildup at the Q-valley at low-temperature (Fig.\,\ref{fig3}C), and (ii) the transfer of spectral weight from the Q- to $\Gamma$-valley at high temperature (see Fig.\,\ref{fig3}D), due to the increase of the electron-phonon scattering phase space caused by the thermal renormalization of the electronic band structure.

\section*{Discussion}
This TR-ARPES work demonstrates that temperature-induced renormalization of the conduction band in Bi$_2$Te$_3$ plays a critical role in dictating ultrafast electron dynamics. Indeed, as temperature rises, the downward shift of the conduction band minimum at $\Gamma$ enhances the spectral overlap between the Q- and $\Gamma$-valleys. This modification opens a novel scattering channel that facilitates intervalley electron-phonon scattering, as evidenced by the marked transfer of spectral weight from Q to $\Gamma$ valleys. The experimental trends are supported by kinetic Monte Carlo simulations, which well reproduce the evolution of the photoemission intensity.  
Our findings suggest that the long-lasting intensity buildup does not necessarily involve the emergence of a spatially indirect exciton, but its formation 
is well captured by the temperature-induced change in the intervalley scattering efficiency within the conduction band. 
Ultimately, our work provides a comprehensive understanding of the intricate mechanisms governing the out-of-equilibrium response of topological insulators, as well as essential insights that could drive the rational design of advanced TI-based quantum and thermoelectric devices, where precise control of out-of-equilibrium electron dynamics over a wide temperature range is paramount.

\section*{Methods}
\subsection*{Samples}
The Bi$_2$Te$_3$ crystals were grown from the elements in quartz ampoules using the Bridgman method, as described elsewhere \cite{Michiardi2014}.

\subsection*{TR-ARPES}
TR-ARPES experiments were performed at the TR-ARPES endstation of the Advanced Laser Light Source (ALLS) laboratory \cite{Longa2024}. The sample was photoexcited with 100 fs, p-polarized, mid-IR pulses (300\,meV photon energy), and the photoemission process was elicited by 6\,eV, s-polarized, pulses. Photoelectrons were detected with an ASTRAIOS 190 hemispherical analyzer, and the overall energy and temporal resolutions of the TR-ARPES endstation were set to 15\,meV and 300\,fs, respectively. Throughout this work, the incident fluence of the pump pulse was set to $\sim$21\,$\mu$J/cm$^2$. Experiments were performed with a bias of -10\,V , to extend the angular acceptance for photoelectrons \cite{Gauthier2021}. 

Complementary TR-ARPES work was also done at the Quantum Matter Institute the UBC-Moore Centre for Ultrafast Quantum Matter, using a Scienta DA30L electron analyzer, 6.2\,eV probe and 1.55\,eV pump photons. The sample temperature was 10\,K (see Fig.\,\ref{fig:Fig_SI4}).

\subsection*{KMC calculations}
In KMC calculations, the bulk CB was reproduced by a uniform grid of states that qualitatively captures the momentum-integrated density of states of a 3D system. Considering the absorption or emission of a phonon, the transition matrix of the system, $W_{i,j}$, was obtained using the Fermi's golden rule: 
\begin{align*}
    W_{i,j} = \frac{2\pi}{\hbar} g_0^{2} \{ &[n(\omega _{i,j} )+1]  \delta (E_j-E_i+\hbar \omega_{i,j} )
    + n(\omega _{i,j} )\delta (E_j-E_i-\hbar \omega _{i,j} )] \},
\end{align*}
where $n(\omega )=(e^{\frac{\hbar \omega }{k_B T}}-1)^{-1}$ is the Bose-Einstein distribution, $\omega_{i,j} (q) \propto 1+\alpha \cdot \cos{(\pi q)}$ is the phonon dispersion evaluated at the scattering vector $q=k_i-k_j$, and $g_0$  is the electron-phonon matrix element (assumed constant for simplicity).
Starting from a Gaussian distribution of electrons centred at 0.4\,eV, we evaluate the electron occupancy as a function of time, $f_i (t)$, by solving the differential Eq.\,\ref{eq:rate}. Each step in the KMC simulation is weighted on the total transition rate
\(\Delta t =\frac{ - \ln r}{ \sum_{i,j} W_{i,j}},\)
where $ r\in(0,1]$ is a randomly sampled value.

Given the experimental evidence for intervalley scattering, we use an optical phonon dispersion centred at $\sim4$\,meV with a bandwidth of $\pm\, 3$\,meV \cite{li2015thermal}. 
We also assume a constant electron-phonon matrix element for transitions within the CB and the TSS (same spin-polarized branch), and we set $g_0^{CB\rightarrow TSS}=g_0/2$ to account for the spin polarization of the TSS.
However, note that the particular choice of the phonon dispersion, \textit{i.e.} optical or acoustic modes, and the strength of the electron-phonon matrix element, do not affect the qualitative results of the simulation as long as the (i) phonon emission dominates over phonon absorption, and (ii) phonon population differs from zero in the region of interest ($0<q<0.25\,\text{\AA}^{-1}$) \cite{Sobota_2014}. Within our simple KMC calculations, electron relaxation dynamics are purely driven by an increase in the electron-phonon scattering phase space. Additionally, since our pump excitation is in a perturbative regime, we neglect any ultrafast renormalization of the electron-phonon matrix element \cite{Zheng2022}.


\clearpage


\renewcommand{\thefigure}{S\arabic{figure}}
\renewcommand{\theequation}{S\arabic{equation}}

\setcounter{figure}{0}
\setcounter{equation}{0}

\newpage
\begin{center}
\section*{Supplementary Material}
\end{center}

\subsection{Symmetry of the intensity buildup}
\begin{figure*}[hb]
    \centering
    \includegraphics[width=1\linewidth]{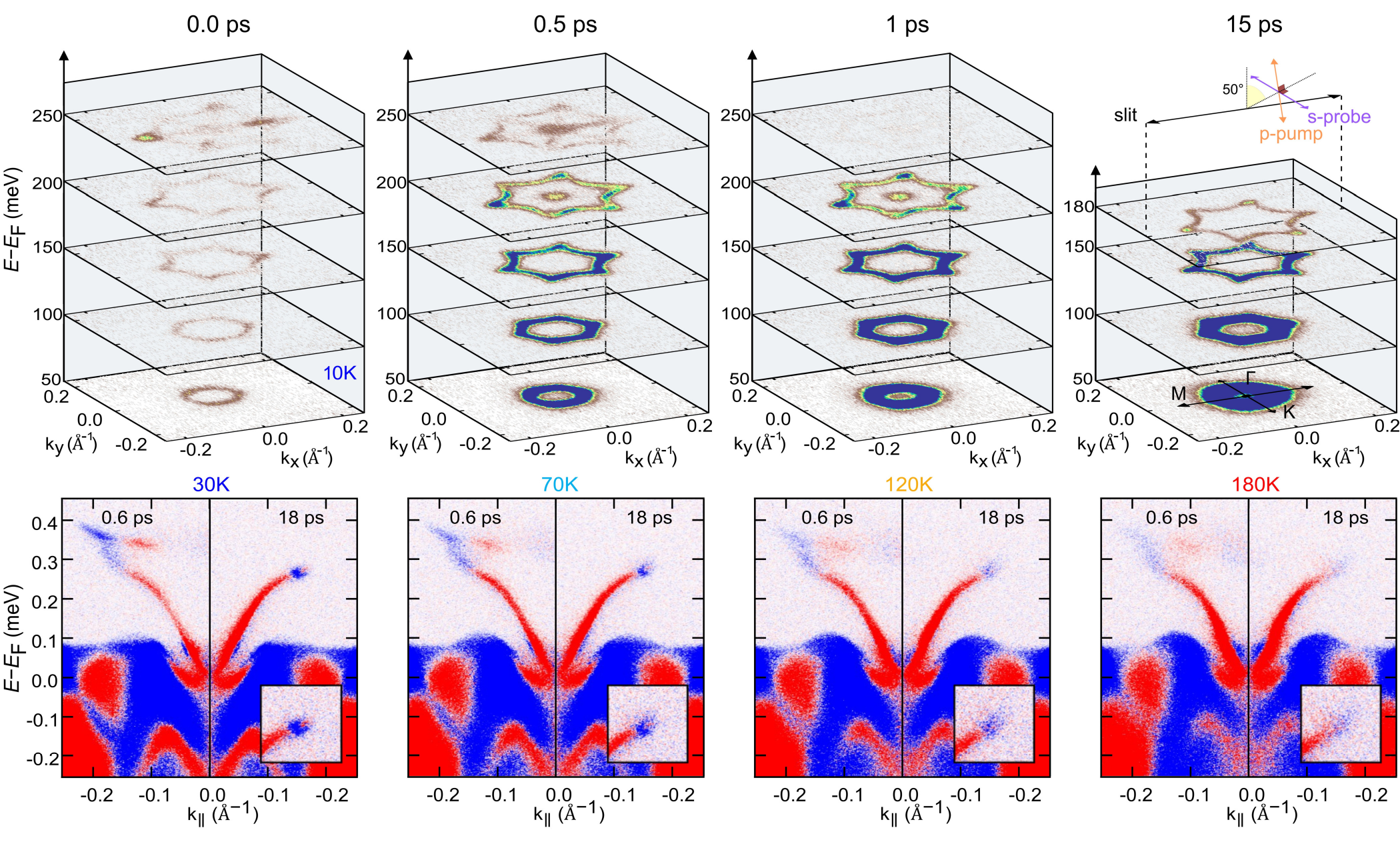}
    \caption{\textbf{Symmetry of the intensity buildup}. Top row: 3D TR-ARPES maps in the ($k_x,k_y,E$) space for different pump-probe delays, from 0 to 15\,ps. The experimental geometry is schematized in the inset above the 15\,ps data. Bottom row: linear dichroism analysis of TR-ARPES data ($I_{LD} \propto I_{k,E} - I_{-k,E}$) for 0.6\,ps and 18\,ps at different temperatures.}
    \label{fig:Fig_SI1}
\end{figure*}

Owing to the use of a hemispherical analyzer with deflector technology in concert with sample biasing, we acquired high-resolution 4D ($k_x,k_y,E, \tau$), TR-ARPES data of p-doped Bi$_2$Te$_3$\cite{Longa2024,Gauthier2021}. Fig.\,\ref{fig:Fig_SI1}, top, displays several iso-energy contour maps, at different central energies ($\pm 20$\,meV integration range) as a function of the pump-probe delay.
Photoexcited electrons populate the conduction band (CB) up to 300\,meV above the Fermi level (E$_\text{F}$), 
and decay and accumulates in states at $\sim$200\,meV above E$_\text{F}$ within 1-2\,ps. These states correspond to the position in the momentum-energy space where the hexagonal warped topological surface state crosses the CB minima and the intensity buildup appears. We note a trigonal intensity pattern for the intensity buildup energy window 
at 0.5\,ps and 1\,ps pump-probe delays, reminiscent of the characteristic bulk trigonal symmetry. At 15\,ps delay, the same trigonal pattern is still present up to 180\,meV above E$_\text{F}$ ($\pm 10$ meV integration range for this central energy), indicating that the long-lasting intensity buildup holds the bulk symmetry.

Further evidence in support of the bulk nature of the intensity buildup is presented in Fig.\,\ref{fig:Fig_SI1}, bottom. A linear dichroism analysis was performed on TR-ARPES data, with the intention of distinguishing between different orbital characters associated of the bulk and surface states. This is possible due to the precise experimental geometry used in our TR-ARPES experiments, \textit{i.e.} s-polarized probe with mirror plane along the $\Gamma$-M direction (time-reversal invariant) \cite{Beaulieu_2021,Min_2019,Cao_2013}. 
Linear dichroism maps were generated by computing $I_{LD} \propto I_{k,E} - I_{-k,E}$. 
Fig.\,\ref{fig:Fig_SI1}, bottom, displays the linear dichroism maps from 30\,K to 180\,K, at 0.6\,ps (left) and 18\,ps (right) pump-probe delays.
The distinct red (positive) and blue (negative) contrast is a signature of different orbital contributions \cite{Beaulieu_2021,Min_2019,Cao_2013}. At 0.6\,ps, the TSS exhibits a pronounced positive dichroic signal, while the CB displays a negative dichroic signal. By 18\,ps, the positive dichroic contribution from the TSS intrudes into the negative dichroic region (as highlighted in the insets). This observation suggests that the TSS intersect the bulk CB at the intensity buildup location, where there is coexistence of both surface and bulk states. 

\subsection{Temperature dependent TR-ARPES}
\begin{figure*}[h]
    \centering
    \includegraphics[width=0.98\linewidth]{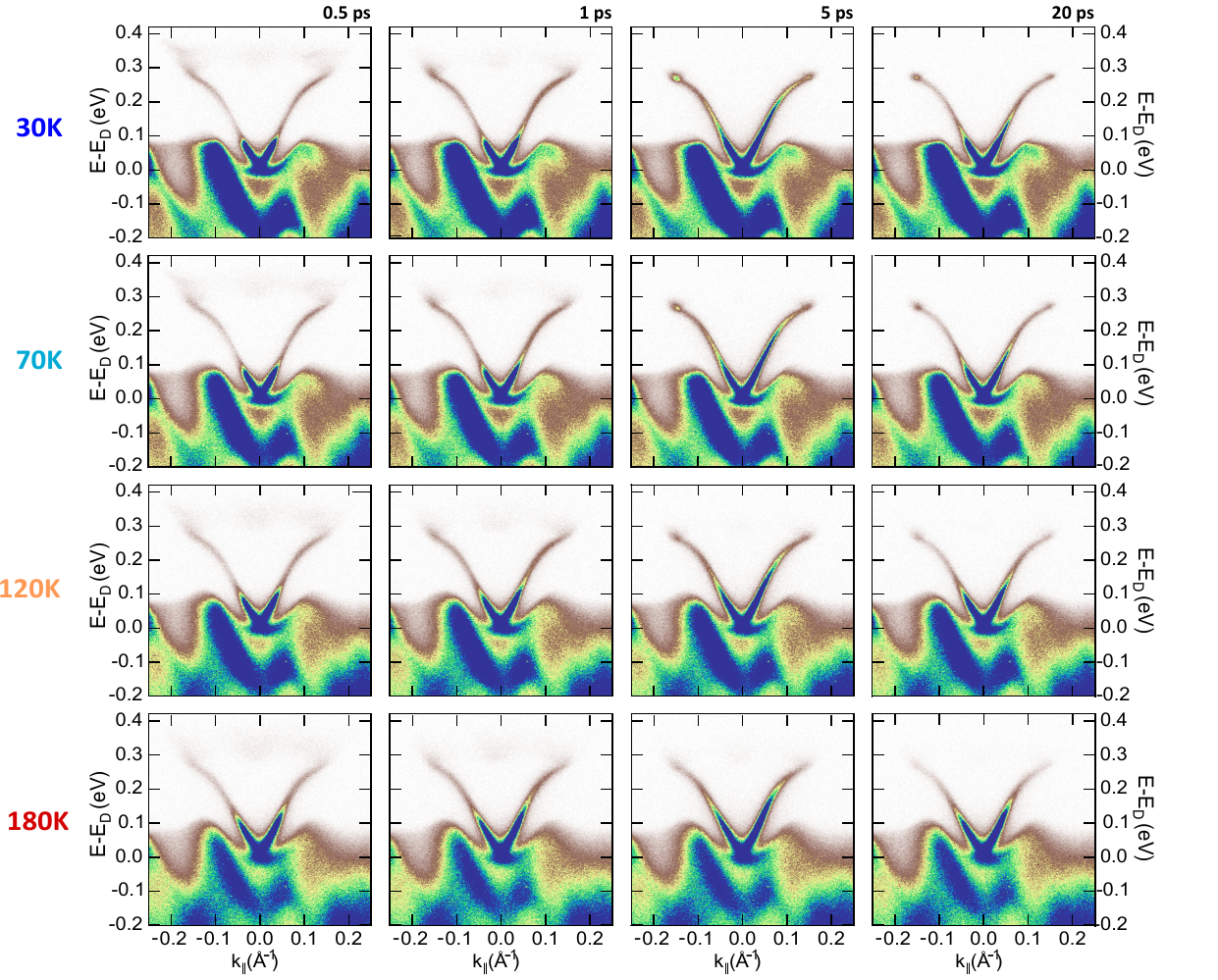}
    \caption{\textbf{Temperature dependent TR-ARPES.} TR-ARPES spectra from 30\,K to 180\,K (from top to bottom) along the $\Gamma$M direction, for different pump probe delays (from 0.5 ps, on the left, to 20 ps, on the right).}
    \label{fig:Fig_SI2}
\end{figure*}

Fig.\,\ref{fig:Fig_SI2} extends the TR-ARPES results of Fig.\,1 in the main text to all the temperatures investigated in this work (30\,K, 70\,K, 120\,K, and 180\,K, top to bottom) at four pump-probe delays (0.5\,ps, 1\,ps, 5\,ps and 20\,ps). The sample displayed minimal n-doping ($\sim$10\,meV) as the temperature increased (as well already discussed elsewhere \cite{Papalazarou_2018}). To facilitate the discussion of our experimental results, the energy axis of the TR-ARPES data is referenced with respect to the Dirac point position. 
This same dataset is used to extract the time-integrated TR-ARPES map of Fig.\,2 in the main text. 

\newpage
\subsection{Electron-phonon toy-model}
\begin{figure*}[h]
    \centering
    \includegraphics[width=0.98\linewidth]{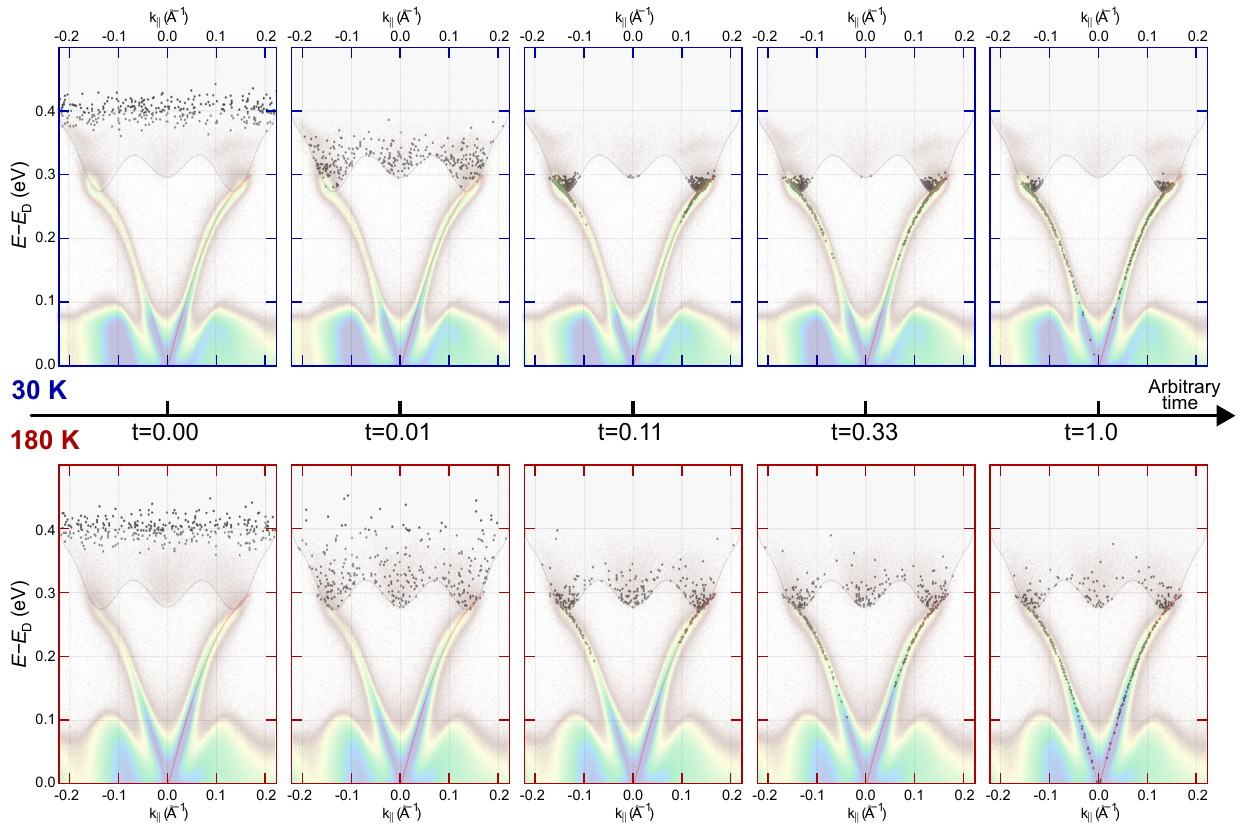}
    \caption{\textbf{Electron-phonon toy-model}. Time-resolved electron occupancies calculated by the KMC toy-model at 30\,K and 180\,K, shown in the top and bottom row, respectively. The time axis is in arbitrary units.}
    \label{fig:Fig_SI3}
\end{figure*}
Fig.\,\ref{fig:Fig_SI3} displays different snapshots of the electron occupancy calculated using the KMC toy-model discussed in the main text (each black circle represents an electron in the energy-momentum space), and using two slightly different band structures of 30\,K and 180\,K (top and bottom row, respectively). The electron-phonon matrix element for intervalley transitions within the CB is assumed constant, $g_0$. Regarding electron scattering from the CB to the TSS, we set the electron-phonon matrix element to $g_0/2$ and, owing to its protection against back-scattering events, the scattering between different branches of the TSS is neglected for simplicity.
As the arbitrary time of the simulation evolves, the continuous emission of phonons favors the decay of electrons in lower energy states, down to the minima of the CB where the TSS crosses the CB. Here, we report a strong bottleneck effect. Indeed, (i) the reduced density of states of TSS compared to that of the CB, and (ii) the lack of TSS at the $\Gamma$ valley, favour the formation of the characteristic intensity buildup (at high momenta) at low temperatures. 
However, at high temperature, the higher phonon population due to Bose-Einstein distribution, in combination with the down-shift of the CB minimum at $\Gamma$, lead to an enhancement of electron scattering towards the $\Gamma$-valley, thus reducing the occupation of the Q-valleys. 

\subsection{Energy position of the intensity buildup}
\begin{figure}[h]
    \centering
    \includegraphics[width=0.99\linewidth]{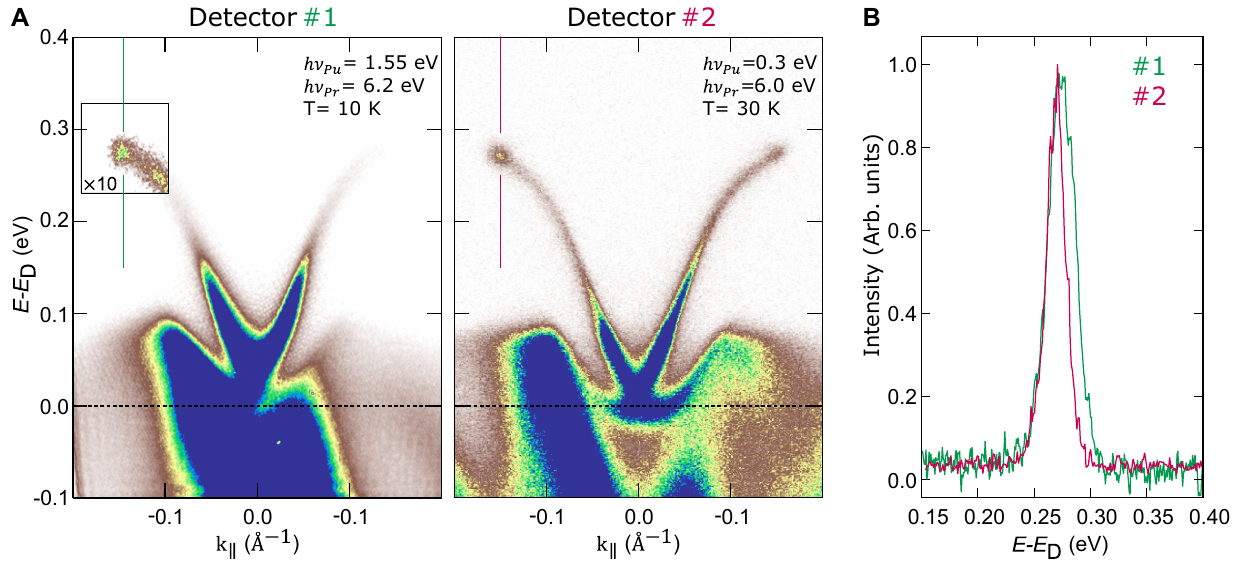}
    \caption{\textbf{Energy position of the intensity buildup.} (A) TR-ARPES spectra at 20\(\pm2\) ps acquired on Bi$_2$Te$_3$ samples using different experimental apparatus. Despite the difference in both pump-probe energies and electron detectors (Scienta DA30L on the left, and SPECS ASTRAIOS 190 on the right), the energy position of the intensity build up is comparable (within 10 meV), as indicated by the EDCs plotted on panel (B).}
    \label{fig:Fig_SI4}
\end{figure}

We compare the results of two independent TR-ARPES experiments performed on p-doped Bi$_2$Te$_3$\ samples from the same batch, but using different detectors (namely, SPECS ASTRAIOS 190 and Scienta DA30L), probe and pump photon energies, and base temperatures.
As shown in Fig.\,\ref{fig:Fig_SI4}A, the intensity buildup appears consistently at \(272\pm5\,\mathrm{meV}\), as highlighted by the EDCs in Fig.\,\ref{fig:Fig_SI4}B. We note that this value corresponds to an energy shift of more than \(\sim30\,\mathrm{meV}\) with respect to the value reported in Ref.\,\onlinecite{Mori2023}, where the same spectroscopic feature is more than \(0.30\,\mathrm{eV}\) (at 20\,ps) above the Dirac point.

\newpage
\begin{acknowledgments}
We thank the ALLS technical team for their support in the laboratory.
The work at ALLS was supported by the Canada Foundation for Innovation (CFI) -- Major Science Initiatives. 
This research was undertaken thanks in part to funding from the Max Planck–UBC–UTokyo Centre for Quantum Materials and the Canada First Research Excellence Fund, Quantum Materials and Future Technologies.
This project is also funded by
the Natural Sciences and Engineering Research Council of Canada, the Canada Research Chairs Program (F.B., A.D.), the CFI, the Department of National Defence (DND), the Fonds de recherche du Qu\'{e}bec --- Nature et Technologies, the Minist\`{e}re de l'\'{E}conomie, de l'Innovation et de l'\'{E}nergie --- Qu\'{e}bec, PRIMA Qu\'{e}bec, the British Columbia Knowledge Development Fund (BCKDF), the CIFAR Quantum Materials Program (A.D.), the Gordon and Betty Moore Foundation’s EPiQS Initiative, grant GBMF12761 (F.B., F.L.) and grant GBMF4779 (A.D.). F.G. acknowledges support from the QuantEmX grant from ICAM and the Gordon and Betty Moore Foundation through Grant GBMF9616. P.H. acknowledges the support of the Independent Research Fund Denmark (grant 4258-00002B) and the Novo Nordisk Foundation (grant NFF23OC0085585).
\end{acknowledgments}

\newpage
\begin{center}
\section*{References}
\end{center}


\begin{thebibliography}{54}%
\makeatletter
\providecommand \@ifxundefined [1]{%
 \@ifx{#1\undefined}
}%
\providecommand \@ifnum [1]{%
 \ifnum #1\expandafter \@firstoftwo
 \else \expandafter \@secondoftwo
 \fi
}%
\providecommand \@ifx [1]{%
 \ifx #1\expandafter \@firstoftwo
 \else \expandafter \@secondoftwo
 \fi
}%
\providecommand \natexlab [1]{#1}%
\providecommand \enquote  [1]{``#1''}%
\providecommand \bibnamefont  [1]{#1}%
\providecommand \bibfnamefont [1]{#1}%
\providecommand \citenamefont [1]{#1}%
\providecommand \href@noop [0]{\@secondoftwo}%
\providecommand \href [0]{\begingroup \@sanitize@url \@href}%
\providecommand \@href[1]{\@@startlink{#1}\@@href}%
\providecommand \@@href[1]{\endgroup#1\@@endlink}%
\providecommand \@sanitize@url [0]{\catcode `\\12\catcode `\$12\catcode `\&12\catcode `\#12\catcode `\^12\catcode `\_12\catcode `\%12\relax}%
\providecommand \@@startlink[1]{}%
\providecommand \@@endlink[0]{}%
\providecommand \url  [0]{\begingroup\@sanitize@url \@url }%
\providecommand \@url [1]{\endgroup\@href {#1}{\urlprefix }}%
\providecommand \urlprefix  [0]{URL }%
\providecommand \Eprint [0]{\href }%
\providecommand \doibase [0]{http://dx.doi.org/}%
\providecommand \selectlanguage [0]{\@gobble}%
\providecommand \bibinfo  [0]{\@secondoftwo}%
\providecommand \bibfield  [0]{\@secondoftwo}%
\providecommand \translation [1]{[#1]}%
\providecommand \BibitemOpen [0]{}%
\providecommand \bibitemStop [0]{}%
\providecommand \bibitemNoStop [0]{.\EOS\space}%
\providecommand \EOS [0]{\spacefactor3000\relax}%
\providecommand \BibitemShut  [1]{\csname bibitem#1\endcsname}%
\let\auto@bib@innerbib\@empty
\bibitem [{\citenamefont {Fu}\ \emph {et~al.}(2007)\citenamefont {Fu}, \citenamefont {Kane},\ and\ \citenamefont {Mele}}]{Fu_2007}%
  \BibitemOpen
  \bibfield  {author} {\bibinfo {author} {\bibfnamefont {L.}~\bibnamefont {Fu}}, \bibinfo {author} {\bibfnamefont {C.~L.}\ \bibnamefont {Kane}}, \ and\ \bibinfo {author} {\bibfnamefont {E.~J.}\ \bibnamefont {Mele}},\ }\href {\doibase 10.1103/physrevlett.98.106803} {\bibfield  {journal} {\bibinfo  {journal} {Physical Review Letters}\ }\textbf {\bibinfo {volume} {98}},\ \bibinfo {pages} {106803} (\bibinfo {year} {2007})}\BibitemShut {NoStop}%
\bibitem [{\citenamefont {Moore}(2010)}]{Moore_2010}%
  \BibitemOpen
  \bibfield  {author} {\bibinfo {author} {\bibfnamefont {J.~E.}\ \bibnamefont {Moore}},\ }\href {\doibase 10.1038/nature08916} {\bibfield  {journal} {\bibinfo  {journal} {Nature}\ }\textbf {\bibinfo {volume} {464}},\ \bibinfo {pages} {194–198} (\bibinfo {year} {2010})}\BibitemShut {NoStop}%
\bibitem [{\citenamefont {Chen}\ \emph {et~al.}(2009)\citenamefont {Chen}, \citenamefont {Analytis}, \citenamefont {Chu}, \citenamefont {Liu}, \citenamefont {Mo}, \citenamefont {Qi}, \citenamefont {Zhang}, \citenamefont {Lu}, \citenamefont {Dai}, \citenamefont {Fang}, \citenamefont {Zhang}, \citenamefont {Fisher}, \citenamefont {Hussain},\ and\ \citenamefont {Shen}}]{Chen_2009}%
  \BibitemOpen
  \bibfield  {author} {\bibinfo {author} {\bibfnamefont {Y.~L.}\ \bibnamefont {Chen}}, \bibinfo {author} {\bibfnamefont {J.~G.}\ \bibnamefont {Analytis}}, \bibinfo {author} {\bibfnamefont {J.-H.}\ \bibnamefont {Chu}}, \bibinfo {author} {\bibfnamefont {Z.~K.}\ \bibnamefont {Liu}}, \bibinfo {author} {\bibfnamefont {S.-K.}\ \bibnamefont {Mo}}, \bibinfo {author} {\bibfnamefont {X.~L.}\ \bibnamefont {Qi}}, \bibinfo {author} {\bibfnamefont {H.~J.}\ \bibnamefont {Zhang}}, \bibinfo {author} {\bibfnamefont {D.~H.}\ \bibnamefont {Lu}}, \bibinfo {author} {\bibfnamefont {X.}~\bibnamefont {Dai}}, \bibinfo {author} {\bibfnamefont {Z.}~\bibnamefont {Fang}}, \bibinfo {author} {\bibfnamefont {S.~C.}\ \bibnamefont {Zhang}}, \bibinfo {author} {\bibfnamefont {I.~R.}\ \bibnamefont {Fisher}}, \bibinfo {author} {\bibfnamefont {Z.}~\bibnamefont {Hussain}}, \ and\ \bibinfo {author} {\bibfnamefont {Z.-X.}\ \bibnamefont {Shen}},\ }\href {\doibase 10.1126/science.1173034} {\bibfield  {journal} {\bibinfo  {journal} {Science}\ }\textbf
  {\bibinfo {volume} {325}},\ \bibinfo {pages} {178–181} (\bibinfo {year} {2009})}\BibitemShut {NoStop}%
\bibitem [{\citenamefont {Hasan}\ and\ \citenamefont {Kane}(2010)}]{Hasan_2010}%
  \BibitemOpen
  \bibfield  {author} {\bibinfo {author} {\bibfnamefont {M.~Z.}\ \bibnamefont {Hasan}}\ and\ \bibinfo {author} {\bibfnamefont {C.~L.}\ \bibnamefont {Kane}},\ }\href {\doibase 10.1103/revmodphys.82.3045} {\bibfield  {journal} {\bibinfo  {journal} {Reviews of Modern Physics}\ }\textbf {\bibinfo {volume} {82}},\ \bibinfo {pages} {3045–3067} (\bibinfo {year} {2010})}\BibitemShut {NoStop}%
\bibitem [{\citenamefont {Ferreira}\ and\ \citenamefont {Loss}(2013)}]{Ferreira_2013}%
  \BibitemOpen
  \bibfield  {author} {\bibinfo {author} {\bibfnamefont {G.~J.}\ \bibnamefont {Ferreira}}\ and\ \bibinfo {author} {\bibfnamefont {D.}~\bibnamefont {Loss}},\ }\href {\doibase 10.1103/physrevlett.111.106802} {\bibfield  {journal} {\bibinfo  {journal} {Physical Review Letters}\ }\textbf {\bibinfo {volume} {111}},\ \bibinfo {pages} {106802} (\bibinfo {year} {2013})}\BibitemShut {NoStop}%
\bibitem [{\citenamefont {Jin}\ \emph {et~al.}(2023)\citenamefont {Jin}, \citenamefont {Jiang}, \citenamefont {Sethi},\ and\ \citenamefont {Liu}}]{Jin_2023}%
  \BibitemOpen
  \bibfield  {author} {\bibinfo {author} {\bibfnamefont {K.-H.}\ \bibnamefont {Jin}}, \bibinfo {author} {\bibfnamefont {W.}~\bibnamefont {Jiang}}, \bibinfo {author} {\bibfnamefont {G.}~\bibnamefont {Sethi}}, \ and\ \bibinfo {author} {\bibfnamefont {F.}~\bibnamefont {Liu}},\ }\href {\doibase 10.1039/d3nr01288c} {\bibfield  {journal} {\bibinfo  {journal} {Nanoscale}\ }\textbf {\bibinfo {volume} {15}},\ \bibinfo {pages} {12787–12817} (\bibinfo {year} {2023})}\BibitemShut {NoStop}%
\bibitem [{\citenamefont {Tao}\ \emph {et~al.}(2020)\citenamefont {Tao}, \citenamefont {Jiang}, \citenamefont {Hao}, \citenamefont {Zheng}, \citenamefont {Zhang},\ and\ \citenamefont {Zeng}}]{tao2020pure}%
  \BibitemOpen
  \bibfield  {author} {\bibinfo {author} {\bibfnamefont {X.}~\bibnamefont {Tao}}, \bibinfo {author} {\bibfnamefont {P.}~\bibnamefont {Jiang}}, \bibinfo {author} {\bibfnamefont {H.}~\bibnamefont {Hao}}, \bibinfo {author} {\bibfnamefont {X.}~\bibnamefont {Zheng}}, \bibinfo {author} {\bibfnamefont {L.}~\bibnamefont {Zhang}}, \ and\ \bibinfo {author} {\bibfnamefont {Z.}~\bibnamefont {Zeng}},\ }\href@noop {} {\bibfield  {journal} {\bibinfo  {journal} {Physical Review B}\ }\textbf {\bibinfo {volume} {102}},\ \bibinfo {pages} {081402} (\bibinfo {year} {2020})}\BibitemShut {NoStop}%
\bibitem [{\citenamefont {Haldane}(2017)}]{Haldane_2017}%
  \BibitemOpen
  \bibfield  {author} {\bibinfo {author} {\bibfnamefont {F.~D.~M.}\ \bibnamefont {Haldane}},\ }\href {\doibase 10.1103/revmodphys.89.040502} {\bibfield  {journal} {\bibinfo  {journal} {Reviews of Modern Physics}\ }\textbf {\bibinfo {volume} {89}},\ \bibinfo {pages} {040502} (\bibinfo {year} {2017})}\BibitemShut {NoStop}%
\bibitem [{\citenamefont {Hsieh}\ \emph {et~al.}(2009)\citenamefont {Hsieh}, \citenamefont {Xia}, \citenamefont {Wray}, \citenamefont {Qian}, \citenamefont {Pal}, \citenamefont {Dil}, \citenamefont {Osterwalder}, \citenamefont {Meier}, \citenamefont {Bihlmayer}, \citenamefont {Kane}, \citenamefont {Hor}, \citenamefont {Cava},\ and\ \citenamefont {Hasan}}]{Hsieh2009ObservationInsulators}%
  \BibitemOpen
  \bibfield  {author} {\bibinfo {author} {\bibfnamefont {D.}~\bibnamefont {Hsieh}}, \bibinfo {author} {\bibfnamefont {Y.}~\bibnamefont {Xia}}, \bibinfo {author} {\bibfnamefont {L.}~\bibnamefont {Wray}}, \bibinfo {author} {\bibfnamefont {D.}~\bibnamefont {Qian}}, \bibinfo {author} {\bibfnamefont {A.}~\bibnamefont {Pal}}, \bibinfo {author} {\bibfnamefont {J.~H.}\ \bibnamefont {Dil}}, \bibinfo {author} {\bibfnamefont {J.}~\bibnamefont {Osterwalder}}, \bibinfo {author} {\bibfnamefont {F.}~\bibnamefont {Meier}}, \bibinfo {author} {\bibfnamefont {G.}~\bibnamefont {Bihlmayer}}, \bibinfo {author} {\bibfnamefont {C.~L.}\ \bibnamefont {Kane}}, \bibinfo {author} {\bibfnamefont {Y.~S.}\ \bibnamefont {Hor}}, \bibinfo {author} {\bibfnamefont {R.~J.}\ \bibnamefont {Cava}}, \ and\ \bibinfo {author} {\bibfnamefont {M.~Z.}\ \bibnamefont {Hasan}},\ }\href {\doibase 10.1126/science.1167733} {\bibfield  {journal} {\bibinfo  {journal} {Science}\ }\textbf {\bibinfo {volume} {323}},\ \bibinfo {pages} {919} (\bibinfo {year}
  {2009})}\BibitemShut {NoStop}%
\bibitem [{\citenamefont {Michiardi}\ \emph {et~al.}(2022)\citenamefont {Michiardi}, \citenamefont {Boschini}, \citenamefont {Kung}, \citenamefont {Na}, \citenamefont {Dufresne}, \citenamefont {Currie}, \citenamefont {Levy}, \citenamefont {Zhdanovich}, \citenamefont {Mills}, \citenamefont {Jones} \emph {et~al.}}]{michiardi2022optical}%
  \BibitemOpen
  \bibfield  {author} {\bibinfo {author} {\bibfnamefont {M.}~\bibnamefont {Michiardi}}, \bibinfo {author} {\bibfnamefont {F.}~\bibnamefont {Boschini}}, \bibinfo {author} {\bibfnamefont {H.-H.}\ \bibnamefont {Kung}}, \bibinfo {author} {\bibfnamefont {M.}~\bibnamefont {Na}}, \bibinfo {author} {\bibfnamefont {S.}~\bibnamefont {Dufresne}}, \bibinfo {author} {\bibfnamefont {A.}~\bibnamefont {Currie}}, \bibinfo {author} {\bibfnamefont {G.}~\bibnamefont {Levy}}, \bibinfo {author} {\bibfnamefont {S.}~\bibnamefont {Zhdanovich}}, \bibinfo {author} {\bibfnamefont {A.}~\bibnamefont {Mills}}, \bibinfo {author} {\bibfnamefont {D.}~\bibnamefont {Jones}},  \emph {et~al.},\ }\href@noop {} {\bibfield  {journal} {\bibinfo  {journal} {nature Communications}\ }\textbf {\bibinfo {volume} {13}},\ \bibinfo {pages} {3096} (\bibinfo {year} {2022})}\BibitemShut {NoStop}%
\bibitem [{\citenamefont {Qi}\ and\ \citenamefont {Zhang}(2011)}]{Qi_2011}%
  \BibitemOpen
  \bibfield  {author} {\bibinfo {author} {\bibfnamefont {X.-L.}\ \bibnamefont {Qi}}\ and\ \bibinfo {author} {\bibfnamefont {S.-C.}\ \bibnamefont {Zhang}},\ }\href {\doibase 10.1103/revmodphys.83.1057} {\bibfield  {journal} {\bibinfo  {journal} {Reviews of Modern Physics}\ }\textbf {\bibinfo {volume} {83}},\ \bibinfo {pages} {1057–1110} (\bibinfo {year} {2011})}\BibitemShut {NoStop}%
\bibitem [{\citenamefont {Legg}\ \emph {et~al.}(2021)\citenamefont {Legg}, \citenamefont {Loss},\ and\ \citenamefont {Klinovaja}}]{Legg_2021}%
  \BibitemOpen
  \bibfield  {author} {\bibinfo {author} {\bibfnamefont {H.~F.}\ \bibnamefont {Legg}}, \bibinfo {author} {\bibfnamefont {D.}~\bibnamefont {Loss}}, \ and\ \bibinfo {author} {\bibfnamefont {J.}~\bibnamefont {Klinovaja}},\ }\href {\doibase 10.1103/physrevb.104.165405} {\bibfield  {journal} {\bibinfo  {journal} {Physical Review B}\ }\textbf {\bibinfo {volume} {104}},\ \bibinfo {pages} {165405} (\bibinfo {year} {2021})}\BibitemShut {NoStop}%
\bibitem [{\citenamefont {Paudel}\ and\ \citenamefont {Leuenberger}(2013)}]{Paudel_2013}%
  \BibitemOpen
  \bibfield  {author} {\bibinfo {author} {\bibfnamefont {H.~P.}\ \bibnamefont {Paudel}}\ and\ \bibinfo {author} {\bibfnamefont {M.~N.}\ \bibnamefont {Leuenberger}},\ }\href {\doibase 10.1103/physrevb.88.085316} {\bibfield  {journal} {\bibinfo  {journal} {Physical Review B}\ }\textbf {\bibinfo {volume} {88}},\ \bibinfo {pages} {085316} (\bibinfo {year} {2013})}\BibitemShut {NoStop}%
\bibitem [{\citenamefont {Boschini}\ \emph {et~al.}(2024)\citenamefont {Boschini}, \citenamefont {Zonno},\ and\ \citenamefont {Damascelli}}]{Boschini2024}%
  \BibitemOpen
  \bibfield  {author} {\bibinfo {author} {\bibfnamefont {F.}~\bibnamefont {Boschini}}, \bibinfo {author} {\bibfnamefont {M.}~\bibnamefont {Zonno}}, \ and\ \bibinfo {author} {\bibfnamefont {A.}~\bibnamefont {Damascelli}},\ }\href {\doibase 10.1103/revmodphys.96.015003} {\bibfield  {journal} {\bibinfo  {journal} {Reviews of Modern Physics}\ }\textbf {\bibinfo {volume} {96}},\ \bibinfo {pages} {015003} (\bibinfo {year} {2024})}\BibitemShut {NoStop}%
\bibitem [{\citenamefont {Zonno}\ \emph {et~al.}(2021)\citenamefont {Zonno}, \citenamefont {Boschini},\ and\ \citenamefont {Damascelli}}]{Zonno2021}%
  \BibitemOpen
  \bibfield  {author} {\bibinfo {author} {\bibfnamefont {M.}~\bibnamefont {Zonno}}, \bibinfo {author} {\bibfnamefont {F.}~\bibnamefont {Boschini}}, \ and\ \bibinfo {author} {\bibfnamefont {A.}~\bibnamefont {Damascelli}},\ }\href {\doibase 10.1016/j.elspec.2021.147091} {\bibfield  {journal} {\bibinfo  {journal} {Journal of Electron Spectroscopy and Related Phenomena}\ }\textbf {\bibinfo {volume} {251}},\ \bibinfo {pages} {147091} (\bibinfo {year} {2021})}\BibitemShut {NoStop}%
\bibitem [{\citenamefont {Crepaldi}\ \emph {et~al.}(2013)\citenamefont {Crepaldi}, \citenamefont {Cilento}, \citenamefont {Ressel}, \citenamefont {Cacho}, \citenamefont {Johannsen}, \citenamefont {Zacchigna}, \citenamefont {Berger}, \citenamefont {Bugnon}, \citenamefont {Grazioli}, \citenamefont {Turcu}, \citenamefont {Springate}, \citenamefont {Kern}, \citenamefont {Grioni},\ and\ \citenamefont {Parmigiani}}]{Crepaldi_2013}%
  \BibitemOpen
  \bibfield  {author} {\bibinfo {author} {\bibfnamefont {A.}~\bibnamefont {Crepaldi}}, \bibinfo {author} {\bibfnamefont {F.}~\bibnamefont {Cilento}}, \bibinfo {author} {\bibfnamefont {B.}~\bibnamefont {Ressel}}, \bibinfo {author} {\bibfnamefont {C.}~\bibnamefont {Cacho}}, \bibinfo {author} {\bibfnamefont {J.~C.}\ \bibnamefont {Johannsen}}, \bibinfo {author} {\bibfnamefont {M.}~\bibnamefont {Zacchigna}}, \bibinfo {author} {\bibfnamefont {H.}~\bibnamefont {Berger}}, \bibinfo {author} {\bibfnamefont {P.}~\bibnamefont {Bugnon}}, \bibinfo {author} {\bibfnamefont {C.}~\bibnamefont {Grazioli}}, \bibinfo {author} {\bibfnamefont {I.~C.~E.}\ \bibnamefont {Turcu}}, \bibinfo {author} {\bibfnamefont {E.}~\bibnamefont {Springate}}, \bibinfo {author} {\bibfnamefont {K.}~\bibnamefont {Kern}}, \bibinfo {author} {\bibfnamefont {M.}~\bibnamefont {Grioni}}, \ and\ \bibinfo {author} {\bibfnamefont {F.}~\bibnamefont {Parmigiani}},\ }\href {\doibase 10.1103/physrevb.88.121404} {\bibfield  {journal} {\bibinfo  {journal} {Physical
  Review B}\ }\textbf {\bibinfo {volume} {88}},\ \bibinfo {pages} {121404} (\bibinfo {year} {2013})}\BibitemShut {NoStop}%
\bibitem [{\citenamefont {Sobota}\ \emph {et~al.}(2012)\citenamefont {Sobota}, \citenamefont {Yang}, \citenamefont {Analytis}, \citenamefont {Chen}, \citenamefont {Fisher}, \citenamefont {Kirchmann},\ and\ \citenamefont {Shen}}]{Sobota2012}%
  \BibitemOpen
  \bibfield  {author} {\bibinfo {author} {\bibfnamefont {J.~A.}\ \bibnamefont {Sobota}}, \bibinfo {author} {\bibfnamefont {S.}~\bibnamefont {Yang}}, \bibinfo {author} {\bibfnamefont {J.~G.}\ \bibnamefont {Analytis}}, \bibinfo {author} {\bibfnamefont {Y.~L.}\ \bibnamefont {Chen}}, \bibinfo {author} {\bibfnamefont {I.~R.}\ \bibnamefont {Fisher}}, \bibinfo {author} {\bibfnamefont {P.~S.}\ \bibnamefont {Kirchmann}}, \ and\ \bibinfo {author} {\bibfnamefont {Z.-X.}\ \bibnamefont {Shen}},\ }\href {\doibase 10.1103/PhysRevLett.108.117403} {\bibfield  {journal} {\bibinfo  {journal} {Physical Review Letters}\ }\textbf {\bibinfo {volume} {108}},\ \bibinfo {pages} {117403} (\bibinfo {year} {2012})}\BibitemShut {NoStop}%
\bibitem [{\citenamefont {Sterzi}\ \emph {et~al.}(2017)\citenamefont {Sterzi}, \citenamefont {Manzoni}, \citenamefont {Sbuelz}, \citenamefont {Cilento}, \citenamefont {Zacchigna}, \citenamefont {Bugnon}, \citenamefont {Magrez}, \citenamefont {Berger}, \citenamefont {Crepaldi},\ and\ \citenamefont {Parmigiani}}]{Sterzi_2017}%
  \BibitemOpen
  \bibfield  {author} {\bibinfo {author} {\bibfnamefont {A.}~\bibnamefont {Sterzi}}, \bibinfo {author} {\bibfnamefont {G.}~\bibnamefont {Manzoni}}, \bibinfo {author} {\bibfnamefont {L.}~\bibnamefont {Sbuelz}}, \bibinfo {author} {\bibfnamefont {F.}~\bibnamefont {Cilento}}, \bibinfo {author} {\bibfnamefont {M.}~\bibnamefont {Zacchigna}}, \bibinfo {author} {\bibfnamefont {P.}~\bibnamefont {Bugnon}}, \bibinfo {author} {\bibfnamefont {A.}~\bibnamefont {Magrez}}, \bibinfo {author} {\bibfnamefont {H.}~\bibnamefont {Berger}}, \bibinfo {author} {\bibfnamefont {A.}~\bibnamefont {Crepaldi}}, \ and\ \bibinfo {author} {\bibfnamefont {F.}~\bibnamefont {Parmigiani}},\ }\href {\doibase 10.1103/physrevb.95.115431} {\bibfield  {journal} {\bibinfo  {journal} {Physical Review B}\ }\textbf {\bibinfo {volume} {95}},\ \bibinfo {pages} {115431} (\bibinfo {year} {2017})}\BibitemShut {NoStop}%
\bibitem [{\citenamefont {Neupane}\ \emph {et~al.}(2015)\citenamefont {Neupane}, \citenamefont {Xu}, \citenamefont {Ishida}, \citenamefont {Jia}, \citenamefont {Fregoso}, \citenamefont {Liu}, \citenamefont {Belopolski}, \citenamefont {Bian}, \citenamefont {Alidoust}, \citenamefont {Durakiewicz}, \citenamefont {Galitski}, \citenamefont {Shin}, \citenamefont {Cava},\ and\ \citenamefont {Hasan}}]{Neupane2015}%
  \BibitemOpen
  \bibfield  {author} {\bibinfo {author} {\bibfnamefont {M.}~\bibnamefont {Neupane}}, \bibinfo {author} {\bibfnamefont {S.-Y.}\ \bibnamefont {Xu}}, \bibinfo {author} {\bibfnamefont {Y.}~\bibnamefont {Ishida}}, \bibinfo {author} {\bibfnamefont {S.}~\bibnamefont {Jia}}, \bibinfo {author} {\bibfnamefont {B.~M.}\ \bibnamefont {Fregoso}}, \bibinfo {author} {\bibfnamefont {C.}~\bibnamefont {Liu}}, \bibinfo {author} {\bibfnamefont {I.}~\bibnamefont {Belopolski}}, \bibinfo {author} {\bibfnamefont {G.}~\bibnamefont {Bian}}, \bibinfo {author} {\bibfnamefont {N.}~\bibnamefont {Alidoust}}, \bibinfo {author} {\bibfnamefont {T.}~\bibnamefont {Durakiewicz}}, \bibinfo {author} {\bibfnamefont {V.}~\bibnamefont {Galitski}}, \bibinfo {author} {\bibfnamefont {S.}~\bibnamefont {Shin}}, \bibinfo {author} {\bibfnamefont {R.~J.}\ \bibnamefont {Cava}}, \ and\ \bibinfo {author} {\bibfnamefont {M.~Z.}\ \bibnamefont {Hasan}},\ }\href {\doibase 10.1103/PhysRevLett.115.116801} {\bibfield  {journal} {\bibinfo  {journal} {Physical Review
  Letters}\ }\textbf {\bibinfo {volume} {115}},\ \bibinfo {pages} {116801} (\bibinfo {year} {2015})}\BibitemShut {NoStop}%
\bibitem [{\citenamefont {Huang}\ \emph {et~al.}(2023)\citenamefont {Huang}, \citenamefont {Querales-Flores}, \citenamefont {Teitelbaum}, \citenamefont {Cao}, \citenamefont {Henighan}, \citenamefont {Liu}, \citenamefont {Jiang}, \citenamefont {la~Peña}, \citenamefont {Krapivin}, \citenamefont {Haber}, \citenamefont {Sato}, \citenamefont {Chollet}, \citenamefont {Zhu}, \citenamefont {Katayama}, \citenamefont {Power}, \citenamefont {Allen}, \citenamefont {Rotundu}, \citenamefont {Bailey}, \citenamefont {Uher}, \citenamefont {Trigo}, \citenamefont {Kirchmann}, \citenamefont {Éamonn D.~Murray}, \citenamefont {Shen}, \citenamefont {Savić}, \citenamefont {Fahy}, \citenamefont {Sobota},\ and\ \citenamefont {Reis}}]{Huang2023}%
  \BibitemOpen
  \bibfield  {author} {\bibinfo {author} {\bibfnamefont {Y.}~\bibnamefont {Huang}}, \bibinfo {author} {\bibfnamefont {J.~D.}\ \bibnamefont {Querales-Flores}}, \bibinfo {author} {\bibfnamefont {S.~W.}\ \bibnamefont {Teitelbaum}}, \bibinfo {author} {\bibfnamefont {J.}~\bibnamefont {Cao}}, \bibinfo {author} {\bibfnamefont {T.}~\bibnamefont {Henighan}}, \bibinfo {author} {\bibfnamefont {H.}~\bibnamefont {Liu}}, \bibinfo {author} {\bibfnamefont {M.}~\bibnamefont {Jiang}}, \bibinfo {author} {\bibfnamefont {G.~D.}\ \bibnamefont {la~Peña}}, \bibinfo {author} {\bibfnamefont {V.}~\bibnamefont {Krapivin}}, \bibinfo {author} {\bibfnamefont {J.}~\bibnamefont {Haber}}, \bibinfo {author} {\bibfnamefont {T.}~\bibnamefont {Sato}}, \bibinfo {author} {\bibfnamefont {M.}~\bibnamefont {Chollet}}, \bibinfo {author} {\bibfnamefont {D.}~\bibnamefont {Zhu}}, \bibinfo {author} {\bibfnamefont {T.}~\bibnamefont {Katayama}}, \bibinfo {author} {\bibfnamefont {R.}~\bibnamefont {Power}}, \bibinfo {author} {\bibfnamefont {M.}~\bibnamefont
  {Allen}}, \bibinfo {author} {\bibfnamefont {C.~R.}\ \bibnamefont {Rotundu}}, \bibinfo {author} {\bibfnamefont {T.~P.}\ \bibnamefont {Bailey}}, \bibinfo {author} {\bibfnamefont {C.}~\bibnamefont {Uher}}, \bibinfo {author} {\bibfnamefont {M.}~\bibnamefont {Trigo}}, \bibinfo {author} {\bibfnamefont {P.~S.}\ \bibnamefont {Kirchmann}}, \bibinfo {author} {\bibnamefont {Éamonn D.~Murray}}, \bibinfo {author} {\bibfnamefont {Z.-X.}\ \bibnamefont {Shen}}, \bibinfo {author} {\bibfnamefont {I.}~\bibnamefont {Savić}}, \bibinfo {author} {\bibfnamefont {S.}~\bibnamefont {Fahy}}, \bibinfo {author} {\bibfnamefont {J.~A.}\ \bibnamefont {Sobota}}, \ and\ \bibinfo {author} {\bibfnamefont {D.~A.}\ \bibnamefont {Reis}},\ }\href {\doibase 10.1103/PhysRevX.13.041050} {\bibfield  {journal} {\bibinfo  {journal} {Physical Review X}\ }\textbf {\bibinfo {volume} {13}},\ \bibinfo {pages} {041050} (\bibinfo {year} {2023})}\BibitemShut {NoStop}%
\bibitem [{\citenamefont {Sánchez-Barriga}\ \emph {et~al.}(2016)\citenamefont {Sánchez-Barriga}, \citenamefont {Golias}, \citenamefont {Varykhalov}, \citenamefont {Braun}, \citenamefont {Yashina}, \citenamefont {Schumann}, \citenamefont {Minár}, \citenamefont {Ebert}, \citenamefont {Kornilov},\ and\ \citenamefont {Rader}}]{Barriga_2016}%
  \BibitemOpen
  \bibfield  {author} {\bibinfo {author} {\bibfnamefont {J.}~\bibnamefont {Sánchez-Barriga}}, \bibinfo {author} {\bibfnamefont {E.}~\bibnamefont {Golias}}, \bibinfo {author} {\bibfnamefont {A.}~\bibnamefont {Varykhalov}}, \bibinfo {author} {\bibfnamefont {J.}~\bibnamefont {Braun}}, \bibinfo {author} {\bibfnamefont {L.~V.}\ \bibnamefont {Yashina}}, \bibinfo {author} {\bibfnamefont {R.}~\bibnamefont {Schumann}}, \bibinfo {author} {\bibfnamefont {J.}~\bibnamefont {Minár}}, \bibinfo {author} {\bibfnamefont {H.}~\bibnamefont {Ebert}}, \bibinfo {author} {\bibfnamefont {O.}~\bibnamefont {Kornilov}}, \ and\ \bibinfo {author} {\bibfnamefont {O.}~\bibnamefont {Rader}},\ }\href {\doibase 10.1103/physrevb.93.155426} {\bibfield  {journal} {\bibinfo  {journal} {Physical Review B}\ }\textbf {\bibinfo {volume} {93}},\ \bibinfo {pages} {155426} (\bibinfo {year} {2016})}\BibitemShut {NoStop}%
\bibitem [{\citenamefont {Michiardi}\ \emph {et~al.}(2014)\citenamefont {Michiardi}, \citenamefont {Aguilera}, \citenamefont {Bianchi}, \citenamefont {de~Carvalho}, \citenamefont {Ladeira}, \citenamefont {Teixeira}, \citenamefont {Soares}, \citenamefont {Friedrich}, \citenamefont {Blügel},\ and\ \citenamefont {Hofmann}}]{Michiardi2014}%
  \BibitemOpen
  \bibfield  {author} {\bibinfo {author} {\bibfnamefont {M.}~\bibnamefont {Michiardi}}, \bibinfo {author} {\bibfnamefont {I.}~\bibnamefont {Aguilera}}, \bibinfo {author} {\bibfnamefont {M.}~\bibnamefont {Bianchi}}, \bibinfo {author} {\bibfnamefont {V.~E.}\ \bibnamefont {de~Carvalho}}, \bibinfo {author} {\bibfnamefont {L.~O.}\ \bibnamefont {Ladeira}}, \bibinfo {author} {\bibfnamefont {N.~G.}\ \bibnamefont {Teixeira}}, \bibinfo {author} {\bibfnamefont {E.~A.}\ \bibnamefont {Soares}}, \bibinfo {author} {\bibfnamefont {C.}~\bibnamefont {Friedrich}}, \bibinfo {author} {\bibfnamefont {S.}~\bibnamefont {Blügel}}, \ and\ \bibinfo {author} {\bibfnamefont {P.}~\bibnamefont {Hofmann}},\ }\href {\doibase 10.1103/physrevb.90.075105} {\bibfield  {journal} {\bibinfo  {journal} {Physical Review B}\ }\textbf {\bibinfo {volume} {90}},\ \bibinfo {pages} {075105} (\bibinfo {year} {2014})}\BibitemShut {NoStop}%
\bibitem [{\citenamefont {Hajlaoui}\ \emph {et~al.}(2014)\citenamefont {Hajlaoui}, \citenamefont {Papalazarou}, \citenamefont {Mauchain}, \citenamefont {Perfetti}, \citenamefont {Taleb-Ibrahimi}, \citenamefont {Navarin}, \citenamefont {Monteverde}, \citenamefont {Auban-Senzier}, \citenamefont {Pasquier}, \citenamefont {Moisan}, \citenamefont {Boschetto}, \citenamefont {Neupane}, \citenamefont {Hasan}, \citenamefont {Durakiewicz}, \citenamefont {Jiang}, \citenamefont {Xu}, \citenamefont {Miotkowski}, \citenamefont {Chen}, \citenamefont {Jia}, \citenamefont {Ji}, \citenamefont {Cava},\ and\ \citenamefont {Marsi}}]{Hajlaoui2014}%
  \BibitemOpen
  \bibfield  {author} {\bibinfo {author} {\bibfnamefont {M.}~\bibnamefont {Hajlaoui}}, \bibinfo {author} {\bibfnamefont {E.}~\bibnamefont {Papalazarou}}, \bibinfo {author} {\bibfnamefont {J.}~\bibnamefont {Mauchain}}, \bibinfo {author} {\bibfnamefont {L.}~\bibnamefont {Perfetti}}, \bibinfo {author} {\bibfnamefont {A.}~\bibnamefont {Taleb-Ibrahimi}}, \bibinfo {author} {\bibfnamefont {F.}~\bibnamefont {Navarin}}, \bibinfo {author} {\bibfnamefont {M.}~\bibnamefont {Monteverde}}, \bibinfo {author} {\bibfnamefont {P.}~\bibnamefont {Auban-Senzier}}, \bibinfo {author} {\bibfnamefont {C.}~\bibnamefont {Pasquier}}, \bibinfo {author} {\bibfnamefont {N.}~\bibnamefont {Moisan}}, \bibinfo {author} {\bibfnamefont {D.}~\bibnamefont {Boschetto}}, \bibinfo {author} {\bibfnamefont {M.}~\bibnamefont {Neupane}}, \bibinfo {author} {\bibfnamefont {M.}~\bibnamefont {Hasan}}, \bibinfo {author} {\bibfnamefont {T.}~\bibnamefont {Durakiewicz}}, \bibinfo {author} {\bibfnamefont {Z.}~\bibnamefont {Jiang}}, \bibinfo {author}
  {\bibfnamefont {Y.}~\bibnamefont {Xu}}, \bibinfo {author} {\bibfnamefont {I.}~\bibnamefont {Miotkowski}}, \bibinfo {author} {\bibfnamefont {Y.}~\bibnamefont {Chen}}, \bibinfo {author} {\bibfnamefont {S.}~\bibnamefont {Jia}}, \bibinfo {author} {\bibfnamefont {H.}~\bibnamefont {Ji}}, \bibinfo {author} {\bibfnamefont {R.}~\bibnamefont {Cava}}, \ and\ \bibinfo {author} {\bibfnamefont {M.}~\bibnamefont {Marsi}},\ }\href {\doibase 10.1038/ncomms4003} {\bibfield  {journal} {\bibinfo  {journal} {Nature Communications}\ }\textbf {\bibinfo {volume} {5}},\ \bibinfo {pages} {3003} (\bibinfo {year} {2014})}\BibitemShut {NoStop}%
\bibitem [{\citenamefont {Sobota}\ \emph {et~al.}(2014{\natexlab{a}})\citenamefont {Sobota}, \citenamefont {Yang}, \citenamefont {Leuenberger}, \citenamefont {Kemper}, \citenamefont {Analytis}, \citenamefont {Fisher}, \citenamefont {Kirchmann}, \citenamefont {Devereaux},\ and\ \citenamefont {Shen}}]{Sobota2014Distinguishing}%
  \BibitemOpen
  \bibfield  {author} {\bibinfo {author} {\bibfnamefont {J.~A.}\ \bibnamefont {Sobota}}, \bibinfo {author} {\bibfnamefont {S.-L.}\ \bibnamefont {Yang}}, \bibinfo {author} {\bibfnamefont {D.}~\bibnamefont {Leuenberger}}, \bibinfo {author} {\bibfnamefont {A.~F.}\ \bibnamefont {Kemper}}, \bibinfo {author} {\bibfnamefont {J.~G.}\ \bibnamefont {Analytis}}, \bibinfo {author} {\bibfnamefont {I.~R.}\ \bibnamefont {Fisher}}, \bibinfo {author} {\bibfnamefont {P.~S.}\ \bibnamefont {Kirchmann}}, \bibinfo {author} {\bibfnamefont {T.~P.}\ \bibnamefont {Devereaux}}, \ and\ \bibinfo {author} {\bibfnamefont {Z.-X.}\ \bibnamefont {Shen}},\ }\href {\doibase 10.1103/PhysRevLett.113.157401} {\bibfield  {journal} {\bibinfo  {journal} {Physical Review Letters}\ }\textbf {\bibinfo {volume} {113}},\ \bibinfo {pages} {157401} (\bibinfo {year} {2014}{\natexlab{a}})}\BibitemShut {NoStop}%
\bibitem [{\citenamefont {Sánchez-Barriga}\ \emph {et~al.}(2017)\citenamefont {Sánchez-Barriga}, \citenamefont {Battiato}, \citenamefont {Krivenkov}, \citenamefont {Golias}, \citenamefont {Varykhalov}, \citenamefont {Romualdi}, \citenamefont {Yashina}, \citenamefont {Minár}, \citenamefont {Kornilov}, \citenamefont {Ebert}, \citenamefont {Held},\ and\ \citenamefont {Braun}}]{Barriga2017}%
  \BibitemOpen
  \bibfield  {author} {\bibinfo {author} {\bibfnamefont {J.}~\bibnamefont {Sánchez-Barriga}}, \bibinfo {author} {\bibfnamefont {M.}~\bibnamefont {Battiato}}, \bibinfo {author} {\bibfnamefont {M.}~\bibnamefont {Krivenkov}}, \bibinfo {author} {\bibfnamefont {E.}~\bibnamefont {Golias}}, \bibinfo {author} {\bibfnamefont {A.}~\bibnamefont {Varykhalov}}, \bibinfo {author} {\bibfnamefont {A.}~\bibnamefont {Romualdi}}, \bibinfo {author} {\bibfnamefont {L.~V.}\ \bibnamefont {Yashina}}, \bibinfo {author} {\bibfnamefont {J.}~\bibnamefont {Minár}}, \bibinfo {author} {\bibfnamefont {O.}~\bibnamefont {Kornilov}}, \bibinfo {author} {\bibfnamefont {H.}~\bibnamefont {Ebert}}, \bibinfo {author} {\bibfnamefont {K.}~\bibnamefont {Held}}, \ and\ \bibinfo {author} {\bibfnamefont {J.}~\bibnamefont {Braun}},\ }\href {\doibase 10.1103/PhysRevB.95.125405} {\bibfield  {journal} {\bibinfo  {journal} {Physical Review B}\ }\textbf {\bibinfo {volume} {95}},\ \bibinfo {pages} {125405} (\bibinfo {year} {2017})}\BibitemShut {NoStop}%
\bibitem [{\citenamefont {Hajlaoui}\ \emph {et~al.}(2012)\citenamefont {Hajlaoui}, \citenamefont {Papalazarou}, \citenamefont {Mauchain}, \citenamefont {Lantz}, \citenamefont {Moisan}, \citenamefont {Boschetto}, \citenamefont {Jiang}, \citenamefont {Miotkowski}, \citenamefont {Chen}, \citenamefont {Taleb-Ibrahimi}, \citenamefont {Perfetti},\ and\ \citenamefont {Marsi}}]{Hajlaoui2012}%
  \BibitemOpen
  \bibfield  {author} {\bibinfo {author} {\bibfnamefont {M.}~\bibnamefont {Hajlaoui}}, \bibinfo {author} {\bibfnamefont {E.}~\bibnamefont {Papalazarou}}, \bibinfo {author} {\bibfnamefont {J.}~\bibnamefont {Mauchain}}, \bibinfo {author} {\bibfnamefont {G.}~\bibnamefont {Lantz}}, \bibinfo {author} {\bibfnamefont {N.}~\bibnamefont {Moisan}}, \bibinfo {author} {\bibfnamefont {D.}~\bibnamefont {Boschetto}}, \bibinfo {author} {\bibfnamefont {Z.}~\bibnamefont {Jiang}}, \bibinfo {author} {\bibfnamefont {I.}~\bibnamefont {Miotkowski}}, \bibinfo {author} {\bibfnamefont {Y.~P.}\ \bibnamefont {Chen}}, \bibinfo {author} {\bibfnamefont {A.}~\bibnamefont {Taleb-Ibrahimi}}, \bibinfo {author} {\bibfnamefont {L.}~\bibnamefont {Perfetti}}, \ and\ \bibinfo {author} {\bibfnamefont {M.}~\bibnamefont {Marsi}},\ }\href {\doibase 10.1021/nl301035x} {\bibfield  {journal} {\bibinfo  {journal} {Nano Letters}\ }\textbf {\bibinfo {volume} {12}},\ \bibinfo {pages} {3532} (\bibinfo {year} {2012})}\BibitemShut {NoStop}%
\bibitem [{\citenamefont {Papalazarou}\ \emph {et~al.}(2018)\citenamefont {Papalazarou}, \citenamefont {Khalil}, \citenamefont {Caputo}, \citenamefont {Perfetti}, \citenamefont {Nilforoushan}, \citenamefont {Deng}, \citenamefont {Chen}, \citenamefont {Zhao}, \citenamefont {Taleb-Ibrahimi}, \citenamefont {Konczykowski}, \citenamefont {Hruban}, \citenamefont {Wołoś}, \citenamefont {Materna}, \citenamefont {Krusin-Elbaum},\ and\ \citenamefont {Marsi}}]{Papalazarou_2018}%
  \BibitemOpen
  \bibfield  {author} {\bibinfo {author} {\bibfnamefont {E.}~\bibnamefont {Papalazarou}}, \bibinfo {author} {\bibfnamefont {L.}~\bibnamefont {Khalil}}, \bibinfo {author} {\bibfnamefont {M.}~\bibnamefont {Caputo}}, \bibinfo {author} {\bibfnamefont {L.}~\bibnamefont {Perfetti}}, \bibinfo {author} {\bibfnamefont {N.}~\bibnamefont {Nilforoushan}}, \bibinfo {author} {\bibfnamefont {H.}~\bibnamefont {Deng}}, \bibinfo {author} {\bibfnamefont {Z.}~\bibnamefont {Chen}}, \bibinfo {author} {\bibfnamefont {S.}~\bibnamefont {Zhao}}, \bibinfo {author} {\bibfnamefont {A.}~\bibnamefont {Taleb-Ibrahimi}}, \bibinfo {author} {\bibfnamefont {M.}~\bibnamefont {Konczykowski}}, \bibinfo {author} {\bibfnamefont {A.}~\bibnamefont {Hruban}}, \bibinfo {author} {\bibfnamefont {A.}~\bibnamefont {Wołoś}}, \bibinfo {author} {\bibfnamefont {A.}~\bibnamefont {Materna}}, \bibinfo {author} {\bibfnamefont {L.}~\bibnamefont {Krusin-Elbaum}}, \ and\ \bibinfo {author} {\bibfnamefont {M.}~\bibnamefont {Marsi}},\ }\href {\doibase
  10.1103/physrevmaterials.2.104202} {\bibfield  {journal} {\bibinfo  {journal} {Physical Review Materials}\ }\textbf {\bibinfo {volume} {2}},\ \bibinfo {pages} {104202} (\bibinfo {year} {2018})}\BibitemShut {NoStop}%
\bibitem [{\citenamefont {Soifer}\ \emph {et~al.}(2019)\citenamefont {Soifer}, \citenamefont {Gauthier}, \citenamefont {Kemper}, \citenamefont {Rotundu}, \citenamefont {Yang}, \citenamefont {Xiong}, \citenamefont {Lu}, \citenamefont {Hashimoto}, \citenamefont {Kirchmann}, \citenamefont {Sobota},\ and\ \citenamefont {Shen}}]{Soifer_2019}%
  \BibitemOpen
  \bibfield  {author} {\bibinfo {author} {\bibfnamefont {H.}~\bibnamefont {Soifer}}, \bibinfo {author} {\bibfnamefont {A.}~\bibnamefont {Gauthier}}, \bibinfo {author} {\bibfnamefont {A.~F.}\ \bibnamefont {Kemper}}, \bibinfo {author} {\bibfnamefont {C.~R.}\ \bibnamefont {Rotundu}}, \bibinfo {author} {\bibfnamefont {S.-L.}\ \bibnamefont {Yang}}, \bibinfo {author} {\bibfnamefont {H.}~\bibnamefont {Xiong}}, \bibinfo {author} {\bibfnamefont {D.}~\bibnamefont {Lu}}, \bibinfo {author} {\bibfnamefont {M.}~\bibnamefont {Hashimoto}}, \bibinfo {author} {\bibfnamefont {P.~S.}\ \bibnamefont {Kirchmann}}, \bibinfo {author} {\bibfnamefont {J.~A.}\ \bibnamefont {Sobota}}, \ and\ \bibinfo {author} {\bibfnamefont {Z.-X.}\ \bibnamefont {Shen}},\ }\href {\doibase 10.1103/physrevlett.122.167401} {\bibfield  {journal} {\bibinfo  {journal} {Physical Review Letters}\ }\textbf {\bibinfo {volume} {122}},\ \bibinfo {pages} {167401} (\bibinfo {year} {2019})}\BibitemShut {NoStop}%
\bibitem [{\citenamefont {Chen}\ \emph {et~al.}(2024)\citenamefont {Chen}, \citenamefont {Wang}, \citenamefont {Lin}, \citenamefont {Zhong}, \citenamefont {Zhou}, \citenamefont {Bao}, \citenamefont {Zhang},\ and\ \citenamefont {Zhou}}]{chen2024distinct}%
  \BibitemOpen
  \bibfield  {author} {\bibinfo {author} {\bibfnamefont {W.}~\bibnamefont {Chen}}, \bibinfo {author} {\bibfnamefont {F.}~\bibnamefont {Wang}}, \bibinfo {author} {\bibfnamefont {T.}~\bibnamefont {Lin}}, \bibinfo {author} {\bibfnamefont {H.}~\bibnamefont {Zhong}}, \bibinfo {author} {\bibfnamefont {S.}~\bibnamefont {Zhou}}, \bibinfo {author} {\bibfnamefont {C.}~\bibnamefont {Bao}}, \bibinfo {author} {\bibfnamefont {H.}~\bibnamefont {Zhang}}, \ and\ \bibinfo {author} {\bibfnamefont {S.}~\bibnamefont {Zhou}},\ }\href@noop {} {\bibfield  {journal} {\bibinfo  {journal} {Physical Review B}\ }\textbf {\bibinfo {volume} {110}},\ \bibinfo {pages} {L201116} (\bibinfo {year} {2024})}\BibitemShut {NoStop}%
\bibitem [{\citenamefont {Monserrat}\ and\ \citenamefont {Vanderbilt}(2016)}]{Monserrat2016}%
  \BibitemOpen
  \bibfield  {author} {\bibinfo {author} {\bibfnamefont {B.}~\bibnamefont {Monserrat}}\ and\ \bibinfo {author} {\bibfnamefont {D.}~\bibnamefont {Vanderbilt}},\ }\href {\doibase 10.1103/PhysRevLett.117.226801} {\bibfield  {journal} {\bibinfo  {journal} {Physical Review Letters}\ }\textbf {\bibinfo {volume} {117}},\ \bibinfo {pages} {226801} (\bibinfo {year} {2016})}\BibitemShut {NoStop}%
\bibitem [{\citenamefont {Tanimura}\ \emph {et~al.}(2021)\citenamefont {Tanimura}, \citenamefont {Tanimura},\ and\ \citenamefont {Kanasaki}}]{Tanimura_2021}%
  \BibitemOpen
  \bibfield  {author} {\bibinfo {author} {\bibfnamefont {H.}~\bibnamefont {Tanimura}}, \bibinfo {author} {\bibfnamefont {K.}~\bibnamefont {Tanimura}}, \ and\ \bibinfo {author} {\bibfnamefont {J.}~\bibnamefont {Kanasaki}},\ }\href {\doibase 10.1103/physrevb.104.245201} {\bibfield  {journal} {\bibinfo  {journal} {Physical Review B}\ }\textbf {\bibinfo {volume} {104}},\ \bibinfo {pages} {245201} (\bibinfo {year} {2021})}\BibitemShut {NoStop}%
\bibitem [{\citenamefont {Sjakste}\ \emph {et~al.}(2018)\citenamefont {Sjakste}, \citenamefont {Vast}, \citenamefont {Barbarino}, \citenamefont {Calandra}, \citenamefont {Mauri}, \citenamefont {Kanasaki}, \citenamefont {Tanimura},\ and\ \citenamefont {Tanimura}}]{Sjakste_2018}%
  \BibitemOpen
  \bibfield  {author} {\bibinfo {author} {\bibfnamefont {J.}~\bibnamefont {Sjakste}}, \bibinfo {author} {\bibfnamefont {N.}~\bibnamefont {Vast}}, \bibinfo {author} {\bibfnamefont {G.}~\bibnamefont {Barbarino}}, \bibinfo {author} {\bibfnamefont {M.}~\bibnamefont {Calandra}}, \bibinfo {author} {\bibfnamefont {F.}~\bibnamefont {Mauri}}, \bibinfo {author} {\bibfnamefont {J.}~\bibnamefont {Kanasaki}}, \bibinfo {author} {\bibfnamefont {H.}~\bibnamefont {Tanimura}}, \ and\ \bibinfo {author} {\bibfnamefont {K.}~\bibnamefont {Tanimura}},\ }\href {\doibase 10.1103/physrevb.97.064302} {\bibfield  {journal} {\bibinfo  {journal} {Physical Review B}\ }\textbf {\bibinfo {volume} {97}},\ \bibinfo {pages} {064302} (\bibinfo {year} {2018})}\BibitemShut {NoStop}%
\bibitem [{\citenamefont {Tanimura}\ \emph {et~al.}(2016)\citenamefont {Tanimura}, \citenamefont {Kanasaki}, \citenamefont {Tanimura}, \citenamefont {Sjakste}, \citenamefont {Vast}, \citenamefont {Calandra},\ and\ \citenamefont {Mauri}}]{Tanimura_2016}%
  \BibitemOpen
  \bibfield  {author} {\bibinfo {author} {\bibfnamefont {H.}~\bibnamefont {Tanimura}}, \bibinfo {author} {\bibfnamefont {J.}~\bibnamefont {Kanasaki}}, \bibinfo {author} {\bibfnamefont {K.}~\bibnamefont {Tanimura}}, \bibinfo {author} {\bibfnamefont {J.}~\bibnamefont {Sjakste}}, \bibinfo {author} {\bibfnamefont {N.}~\bibnamefont {Vast}}, \bibinfo {author} {\bibfnamefont {M.}~\bibnamefont {Calandra}}, \ and\ \bibinfo {author} {\bibfnamefont {F.}~\bibnamefont {Mauri}},\ }\href {\doibase 10.1103/physrevb.93.161203} {\bibfield  {journal} {\bibinfo  {journal} {Physical Review B}\ }\textbf {\bibinfo {volume} {93}},\ \bibinfo {pages} {161203} (\bibinfo {year} {2016})}\BibitemShut {NoStop}%
\bibitem [{\citenamefont {Kanasaki}\ \emph {et~al.}(2014)\citenamefont {Kanasaki}, \citenamefont {Tanimura},\ and\ \citenamefont {Tanimura}}]{Kanasaki_2014}%
  \BibitemOpen
  \bibfield  {author} {\bibinfo {author} {\bibfnamefont {J.}~\bibnamefont {Kanasaki}}, \bibinfo {author} {\bibfnamefont {H.}~\bibnamefont {Tanimura}}, \ and\ \bibinfo {author} {\bibfnamefont {K.}~\bibnamefont {Tanimura}},\ }\href {\doibase 10.1103/physrevlett.113.237401} {\bibfield  {journal} {\bibinfo  {journal} {Physical Review Letters}\ }\textbf {\bibinfo {volume} {113}},\ \bibinfo {pages} {237401} (\bibinfo {year} {2014})}\BibitemShut {NoStop}%
\bibitem [{\citenamefont {Tanimura}\ \emph {et~al.}(2015)\citenamefont {Tanimura}, \citenamefont {Kanasaki},\ and\ \citenamefont {Tanimura}}]{Tanimura_2015}%
  \BibitemOpen
  \bibfield  {author} {\bibinfo {author} {\bibfnamefont {H.}~\bibnamefont {Tanimura}}, \bibinfo {author} {\bibfnamefont {J.}~\bibnamefont {Kanasaki}}, \ and\ \bibinfo {author} {\bibfnamefont {K.}~\bibnamefont {Tanimura}},\ }\href {\doibase 10.1103/physrevb.91.045201} {\bibfield  {journal} {\bibinfo  {journal} {Physical Review B}\ }\textbf {\bibinfo {volume} {91}},\ \bibinfo {pages} {045201} (\bibinfo {year} {2015})}\BibitemShut {NoStop}%
\bibitem [{\citenamefont {Ataei}\ and\ \citenamefont {Sadeghi}(2021)}]{ataei2021competitive}%
  \BibitemOpen
  \bibfield  {author} {\bibinfo {author} {\bibfnamefont {S.~S.}\ \bibnamefont {Ataei}}\ and\ \bibinfo {author} {\bibfnamefont {A.}~\bibnamefont {Sadeghi}},\ }\href@noop {} {\bibfield  {journal} {\bibinfo  {journal} {Physical Review B}\ }\textbf {\bibinfo {volume} {104}},\ \bibinfo {pages} {155301} (\bibinfo {year} {2021})}\BibitemShut {NoStop}%
\bibitem [{\citenamefont {Ulstrup}\ \emph {et~al.}(2016)\citenamefont {Ulstrup}, \citenamefont {Cabo}, \citenamefont {Miwa}, \citenamefont {Riley}, \citenamefont {Gr{\o}nborg}, \citenamefont {Johannsen}, \citenamefont {Cacho}, \citenamefont {Alexander}, \citenamefont {Chapman}, \citenamefont {Springate} \emph {et~al.}}]{ulstrup2016ultrafast}%
  \BibitemOpen
  \bibfield  {author} {\bibinfo {author} {\bibfnamefont {S.}~\bibnamefont {Ulstrup}}, \bibinfo {author} {\bibfnamefont {A.~G.}\ \bibnamefont {Cabo}}, \bibinfo {author} {\bibfnamefont {J.~A.}\ \bibnamefont {Miwa}}, \bibinfo {author} {\bibfnamefont {J.~M.}\ \bibnamefont {Riley}}, \bibinfo {author} {\bibfnamefont {S.~S.}\ \bibnamefont {Gr{\o}nborg}}, \bibinfo {author} {\bibfnamefont {J.~C.}\ \bibnamefont {Johannsen}}, \bibinfo {author} {\bibfnamefont {C.}~\bibnamefont {Cacho}}, \bibinfo {author} {\bibfnamefont {O.}~\bibnamefont {Alexander}}, \bibinfo {author} {\bibfnamefont {R.~T.}\ \bibnamefont {Chapman}}, \bibinfo {author} {\bibfnamefont {E.}~\bibnamefont {Springate}},  \emph {et~al.},\ }\href@noop {} {\bibfield  {journal} {\bibinfo  {journal} {ACS nano}\ }\textbf {\bibinfo {volume} {10}},\ \bibinfo {pages} {6315} (\bibinfo {year} {2016})}\BibitemShut {NoStop}%
\bibitem [{\citenamefont {Lee}\ \emph {et~al.}(2021)\citenamefont {Lee}, \citenamefont {Choi}, \citenamefont {Kim}, \citenamefont {Kim},\ and\ \citenamefont {Jung}}]{lee2021direct}%
  \BibitemOpen
  \bibfield  {author} {\bibinfo {author} {\bibfnamefont {D.~H.}\ \bibnamefont {Lee}}, \bibinfo {author} {\bibfnamefont {S.-J.}\ \bibnamefont {Choi}}, \bibinfo {author} {\bibfnamefont {H.}~\bibnamefont {Kim}}, \bibinfo {author} {\bibfnamefont {Y.-S.}\ \bibnamefont {Kim}}, \ and\ \bibinfo {author} {\bibfnamefont {S.}~\bibnamefont {Jung}},\ }\href@noop {} {\bibfield  {journal} {\bibinfo  {journal} {Nature Communications}\ }\textbf {\bibinfo {volume} {12}},\ \bibinfo {pages} {4520} (\bibinfo {year} {2021})}\BibitemShut {NoStop}%
\bibitem [{\citenamefont {Rittweger}\ \emph {et~al.}(2014)\citenamefont {Rittweger}, \citenamefont {Hinsche}, \citenamefont {Zahn},\ and\ \citenamefont {Mertig}}]{Rittweger2014}%
  \BibitemOpen
  \bibfield  {author} {\bibinfo {author} {\bibfnamefont {F.}~\bibnamefont {Rittweger}}, \bibinfo {author} {\bibfnamefont {N.~F.}\ \bibnamefont {Hinsche}}, \bibinfo {author} {\bibfnamefont {P.}~\bibnamefont {Zahn}}, \ and\ \bibinfo {author} {\bibfnamefont {I.}~\bibnamefont {Mertig}},\ }\href {\doibase 10.1103/PhysRevB.89.035439} {\bibfield  {journal} {\bibinfo  {journal} {Physical Review B}\ }\textbf {\bibinfo {volume} {89}},\ \bibinfo {pages} {035439} (\bibinfo {year} {2014})}\BibitemShut {NoStop}%
\bibitem [{\citenamefont {Liang}\ \emph {et~al.}(2016)\citenamefont {Liang}, \citenamefont {Cheng}, \citenamefont {Zhang}, \citenamefont {Liu},\ and\ \citenamefont {Zhang}}]{Liang2016}%
  \BibitemOpen
  \bibfield  {author} {\bibinfo {author} {\bibfnamefont {J.}~\bibnamefont {Liang}}, \bibinfo {author} {\bibfnamefont {L.}~\bibnamefont {Cheng}}, \bibinfo {author} {\bibfnamefont {J.}~\bibnamefont {Zhang}}, \bibinfo {author} {\bibfnamefont {H.}~\bibnamefont {Liu}}, \ and\ \bibinfo {author} {\bibfnamefont {Z.}~\bibnamefont {Zhang}},\ }\href {\doibase 10.1039/C6NR00724D} {\bibfield  {journal} {\bibinfo  {journal} {Nanoscale}\ }\textbf {\bibinfo {volume} {8}},\ \bibinfo {pages} {8855} (\bibinfo {year} {2016})}\BibitemShut {NoStop}%
\bibitem [{\citenamefont {Cao}\ \emph {et~al.}(2023)\citenamefont {Cao}, \citenamefont {Shi}, \citenamefont {Li}, \citenamefont {Hu}, \citenamefont {Chen}, \citenamefont {Liu}, \citenamefont {Lyu}, \citenamefont {MacLeod},\ and\ \citenamefont {Chen}}]{Cao2023}%
  \BibitemOpen
  \bibfield  {author} {\bibinfo {author} {\bibfnamefont {T.}~\bibnamefont {Cao}}, \bibinfo {author} {\bibfnamefont {X.-L.}\ \bibnamefont {Shi}}, \bibinfo {author} {\bibfnamefont {M.}~\bibnamefont {Li}}, \bibinfo {author} {\bibfnamefont {B.}~\bibnamefont {Hu}}, \bibinfo {author} {\bibfnamefont {W.}~\bibnamefont {Chen}}, \bibinfo {author} {\bibfnamefont {W.-D.}\ \bibnamefont {Liu}}, \bibinfo {author} {\bibfnamefont {W.}~\bibnamefont {Lyu}}, \bibinfo {author} {\bibfnamefont {J.}~\bibnamefont {MacLeod}}, \ and\ \bibinfo {author} {\bibfnamefont {Z.-G.}\ \bibnamefont {Chen}},\ }\href {\doibase 10.1016/j.esci.2023.100122} {\bibfield  {journal} {\bibinfo  {journal} {eScience}\ }\textbf {\bibinfo {volume} {3}},\ \bibinfo {pages} {100122} (\bibinfo {year} {2023})}\BibitemShut {NoStop}%
\bibitem [{\citenamefont {Hedayat}\ \emph {et~al.}(2021)\citenamefont {Hedayat}, \citenamefont {Bugini}, \citenamefont {Yi}, \citenamefont {Chen}, \citenamefont {Zhou}, \citenamefont {Cerullo}, \citenamefont {Dallera},\ and\ \citenamefont {Carpene}}]{Hedayat2021}%
  \BibitemOpen
  \bibfield  {author} {\bibinfo {author} {\bibfnamefont {H.}~\bibnamefont {Hedayat}}, \bibinfo {author} {\bibfnamefont {D.}~\bibnamefont {Bugini}}, \bibinfo {author} {\bibfnamefont {H.}~\bibnamefont {Yi}}, \bibinfo {author} {\bibfnamefont {C.}~\bibnamefont {Chen}}, \bibinfo {author} {\bibfnamefont {X.}~\bibnamefont {Zhou}}, \bibinfo {author} {\bibfnamefont {G.}~\bibnamefont {Cerullo}}, \bibinfo {author} {\bibfnamefont {C.}~\bibnamefont {Dallera}}, \ and\ \bibinfo {author} {\bibfnamefont {E.}~\bibnamefont {Carpene}},\ }\href {\doibase 10.1038/s41598-021-83848-z} {\bibfield  {journal} {\bibinfo  {journal} {Scientific Reports}\ }\textbf {\bibinfo {volume} {11}},\ \bibinfo {pages} {4924} (\bibinfo {year} {2021})}\BibitemShut {NoStop}%
\bibitem [{\citenamefont {Cacho}\ \emph {et~al.}(2015)\citenamefont {Cacho}, \citenamefont {Crepaldi}, \citenamefont {Battiato}, \citenamefont {Braun}, \citenamefont {Cilento}, \citenamefont {Zacchigna}, \citenamefont {Richter}, \citenamefont {Heckmann}, \citenamefont {Springate}, \citenamefont {Liu} \emph {et~al.}}]{cacho2015momentum}%
  \BibitemOpen
  \bibfield  {author} {\bibinfo {author} {\bibfnamefont {C.}~\bibnamefont {Cacho}}, \bibinfo {author} {\bibfnamefont {A.}~\bibnamefont {Crepaldi}}, \bibinfo {author} {\bibfnamefont {M.}~\bibnamefont {Battiato}}, \bibinfo {author} {\bibfnamefont {J.}~\bibnamefont {Braun}}, \bibinfo {author} {\bibfnamefont {F.}~\bibnamefont {Cilento}}, \bibinfo {author} {\bibfnamefont {M.}~\bibnamefont {Zacchigna}}, \bibinfo {author} {\bibfnamefont {M.}~\bibnamefont {Richter}}, \bibinfo {author} {\bibfnamefont {O.}~\bibnamefont {Heckmann}}, \bibinfo {author} {\bibfnamefont {E.}~\bibnamefont {Springate}}, \bibinfo {author} {\bibfnamefont {Y.}~\bibnamefont {Liu}},  \emph {et~al.},\ }\href@noop {} {\bibfield  {journal} {\bibinfo  {journal} {Physical review letters}\ }\textbf {\bibinfo {volume} {114}},\ \bibinfo {pages} {097401} (\bibinfo {year} {2015})}\BibitemShut {NoStop}%
\bibitem [{\citenamefont {Jozwiak}\ \emph {et~al.}(2016)\citenamefont {Jozwiak}, \citenamefont {Sobota}, \citenamefont {Gotlieb}, \citenamefont {Kemper}, \citenamefont {Rotundu}, \citenamefont {Birgeneau}, \citenamefont {Hussain}, \citenamefont {Lee}, \citenamefont {Shen},\ and\ \citenamefont {Lanzara}}]{jozwiak2016spin}%
  \BibitemOpen
  \bibfield  {author} {\bibinfo {author} {\bibfnamefont {C.}~\bibnamefont {Jozwiak}}, \bibinfo {author} {\bibfnamefont {J.~A.}\ \bibnamefont {Sobota}}, \bibinfo {author} {\bibfnamefont {K.}~\bibnamefont {Gotlieb}}, \bibinfo {author} {\bibfnamefont {A.~F.}\ \bibnamefont {Kemper}}, \bibinfo {author} {\bibfnamefont {C.~R.}\ \bibnamefont {Rotundu}}, \bibinfo {author} {\bibfnamefont {R.~J.}\ \bibnamefont {Birgeneau}}, \bibinfo {author} {\bibfnamefont {Z.}~\bibnamefont {Hussain}}, \bibinfo {author} {\bibfnamefont {D.-H.}\ \bibnamefont {Lee}}, \bibinfo {author} {\bibfnamefont {Z.-X.}\ \bibnamefont {Shen}}, \ and\ \bibinfo {author} {\bibfnamefont {A.}~\bibnamefont {Lanzara}},\ }\href@noop {} {\bibfield  {journal} {\bibinfo  {journal} {Nature Communications}\ }\textbf {\bibinfo {volume} {7}},\ \bibinfo {pages} {13143} (\bibinfo {year} {2016})}\BibitemShut {NoStop}%
\bibitem [{\citenamefont {Mori}\ \emph {et~al.}(2023)\citenamefont {Mori}, \citenamefont {Ciocys}, \citenamefont {Takasan}, \citenamefont {Ai}, \citenamefont {Currier}, \citenamefont {Morimoto}, \citenamefont {Moore},\ and\ \citenamefont {Lanzara}}]{Mori2023}%
  \BibitemOpen
  \bibfield  {author} {\bibinfo {author} {\bibfnamefont {R.}~\bibnamefont {Mori}}, \bibinfo {author} {\bibfnamefont {S.}~\bibnamefont {Ciocys}}, \bibinfo {author} {\bibfnamefont {K.}~\bibnamefont {Takasan}}, \bibinfo {author} {\bibfnamefont {P.}~\bibnamefont {Ai}}, \bibinfo {author} {\bibfnamefont {K.}~\bibnamefont {Currier}}, \bibinfo {author} {\bibfnamefont {T.}~\bibnamefont {Morimoto}}, \bibinfo {author} {\bibfnamefont {J.~E.}\ \bibnamefont {Moore}}, \ and\ \bibinfo {author} {\bibfnamefont {A.}~\bibnamefont {Lanzara}},\ }\href {\doibase 10.1038/s41586-022-05567-3} {\bibfield  {journal} {\bibinfo  {journal} {Nature}\ }\textbf {\bibinfo {volume} {614}},\ \bibinfo {pages} {249} (\bibinfo {year} {2023})}\BibitemShut {NoStop}%
\bibitem [{\citenamefont {Beaulieu}\ \emph {et~al.}(2021)\citenamefont {Beaulieu}, \citenamefont {Schüler}, \citenamefont {Schusser}, \citenamefont {Dong}, \citenamefont {Pincelli}, \citenamefont {Maklar}, \citenamefont {Neef}, \citenamefont {Reinert}, \citenamefont {Wolf}, \citenamefont {Rettig}, \citenamefont {Minár},\ and\ \citenamefont {Ernstorfer}}]{Beaulieu_2021}%
  \BibitemOpen
  \bibfield  {author} {\bibinfo {author} {\bibfnamefont {S.}~\bibnamefont {Beaulieu}}, \bibinfo {author} {\bibfnamefont {M.}~\bibnamefont {Schüler}}, \bibinfo {author} {\bibfnamefont {J.}~\bibnamefont {Schusser}}, \bibinfo {author} {\bibfnamefont {S.}~\bibnamefont {Dong}}, \bibinfo {author} {\bibfnamefont {T.}~\bibnamefont {Pincelli}}, \bibinfo {author} {\bibfnamefont {J.}~\bibnamefont {Maklar}}, \bibinfo {author} {\bibfnamefont {A.}~\bibnamefont {Neef}}, \bibinfo {author} {\bibfnamefont {F.}~\bibnamefont {Reinert}}, \bibinfo {author} {\bibfnamefont {M.}~\bibnamefont {Wolf}}, \bibinfo {author} {\bibfnamefont {L.}~\bibnamefont {Rettig}}, \bibinfo {author} {\bibfnamefont {J.}~\bibnamefont {Minár}}, \ and\ \bibinfo {author} {\bibfnamefont {R.}~\bibnamefont {Ernstorfer}},\ }\href {\doibase 10.1038/s41535-021-00398-3} {\bibfield  {journal} {\bibinfo  {journal} {npj Quantum Materials}\ }\textbf {\bibinfo {volume} {6}},\ \bibinfo {pages} {93} (\bibinfo {year} {2021})}\BibitemShut {NoStop}%
\bibitem [{\citenamefont {Cao}\ \emph {et~al.}(2013)\citenamefont {Cao}, \citenamefont {Waugh}, \citenamefont {Zhang}, \citenamefont {Luo}, \citenamefont {Wang}, \citenamefont {Reber}, \citenamefont {Mo}, \citenamefont {Xu}, \citenamefont {Yang}, \citenamefont {Schneeloch}, \citenamefont {Gu}, \citenamefont {Brahlek}, \citenamefont {Bansal}, \citenamefont {Oh}, \citenamefont {Zunger},\ and\ \citenamefont {Dessau}}]{Cao_2013}%
  \BibitemOpen
  \bibfield  {author} {\bibinfo {author} {\bibfnamefont {Y.}~\bibnamefont {Cao}}, \bibinfo {author} {\bibfnamefont {J.~A.}\ \bibnamefont {Waugh}}, \bibinfo {author} {\bibfnamefont {X.-W.}\ \bibnamefont {Zhang}}, \bibinfo {author} {\bibfnamefont {J.-W.}\ \bibnamefont {Luo}}, \bibinfo {author} {\bibfnamefont {Q.}~\bibnamefont {Wang}}, \bibinfo {author} {\bibfnamefont {T.~J.}\ \bibnamefont {Reber}}, \bibinfo {author} {\bibfnamefont {S.~K.}\ \bibnamefont {Mo}}, \bibinfo {author} {\bibfnamefont {Z.}~\bibnamefont {Xu}}, \bibinfo {author} {\bibfnamefont {A.}~\bibnamefont {Yang}}, \bibinfo {author} {\bibfnamefont {J.}~\bibnamefont {Schneeloch}}, \bibinfo {author} {\bibfnamefont {G.~D.}\ \bibnamefont {Gu}}, \bibinfo {author} {\bibfnamefont {M.}~\bibnamefont {Brahlek}}, \bibinfo {author} {\bibfnamefont {N.}~\bibnamefont {Bansal}}, \bibinfo {author} {\bibfnamefont {S.}~\bibnamefont {Oh}}, \bibinfo {author} {\bibfnamefont {A.}~\bibnamefont {Zunger}}, \ and\ \bibinfo {author} {\bibfnamefont {D.~S.}\ \bibnamefont
  {Dessau}},\ }\href {\doibase 10.1038/nphys2685} {\bibfield  {journal} {\bibinfo  {journal} {Nature Physics}\ }\textbf {\bibinfo {volume} {9}},\ \bibinfo {pages} {499–504} (\bibinfo {year} {2013})}\BibitemShut {NoStop}%
\bibitem [{\citenamefont {Min}\ \emph {et~al.}(2019)\citenamefont {Min}, \citenamefont {Bentmann}, \citenamefont {Neu}, \citenamefont {Eck}, \citenamefont {Moser}, \citenamefont {Figgemeier}, \citenamefont {Ünzelmann}, \citenamefont {Kissner}, \citenamefont {Lutz}, \citenamefont {Koch}, \citenamefont {Jozwiak}, \citenamefont {Bostwick}, \citenamefont {Rotenberg}, \citenamefont {Thomale}, \citenamefont {Sangiovanni}, \citenamefont {Siegrist}, \citenamefont {Di~Sante},\ and\ \citenamefont {Reinert}}]{Min_2019}%
  \BibitemOpen
  \bibfield  {author} {\bibinfo {author} {\bibfnamefont {C.-H.}\ \bibnamefont {Min}}, \bibinfo {author} {\bibfnamefont {H.}~\bibnamefont {Bentmann}}, \bibinfo {author} {\bibfnamefont {J.~N.}\ \bibnamefont {Neu}}, \bibinfo {author} {\bibfnamefont {P.}~\bibnamefont {Eck}}, \bibinfo {author} {\bibfnamefont {S.}~\bibnamefont {Moser}}, \bibinfo {author} {\bibfnamefont {T.}~\bibnamefont {Figgemeier}}, \bibinfo {author} {\bibfnamefont {M.}~\bibnamefont {Ünzelmann}}, \bibinfo {author} {\bibfnamefont {K.}~\bibnamefont {Kissner}}, \bibinfo {author} {\bibfnamefont {P.}~\bibnamefont {Lutz}}, \bibinfo {author} {\bibfnamefont {R.~J.}\ \bibnamefont {Koch}}, \bibinfo {author} {\bibfnamefont {C.}~\bibnamefont {Jozwiak}}, \bibinfo {author} {\bibfnamefont {A.}~\bibnamefont {Bostwick}}, \bibinfo {author} {\bibfnamefont {E.}~\bibnamefont {Rotenberg}}, \bibinfo {author} {\bibfnamefont {R.}~\bibnamefont {Thomale}}, \bibinfo {author} {\bibfnamefont {G.}~\bibnamefont {Sangiovanni}}, \bibinfo {author} {\bibfnamefont {T.}~\bibnamefont
  {Siegrist}}, \bibinfo {author} {\bibfnamefont {D.}~\bibnamefont {Di~Sante}}, \ and\ \bibinfo {author} {\bibfnamefont {F.}~\bibnamefont {Reinert}},\ }\href {\doibase 10.1103/physrevlett.122.116402} {\bibfield  {journal} {\bibinfo  {journal} {Physical Review Letters}\ }\textbf {\bibinfo {volume} {122}},\ \bibinfo {pages} {116402} (\bibinfo {year} {2019})}\BibitemShut {NoStop}%
\bibitem [{\citenamefont {Sobota}\ \emph {et~al.}(2014{\natexlab{b}})\citenamefont {Sobota}, \citenamefont {Yang}, \citenamefont {Leuenberger}, \citenamefont {Kemper}, \citenamefont {Analytis}, \citenamefont {Fisher}, \citenamefont {Kirchmann}, \citenamefont {Devereaux},\ and\ \citenamefont {Shen}}]{Sobota_2014}%
  \BibitemOpen
  \bibfield  {author} {\bibinfo {author} {\bibfnamefont {J.}~\bibnamefont {Sobota}}, \bibinfo {author} {\bibfnamefont {S.-L.}\ \bibnamefont {Yang}}, \bibinfo {author} {\bibfnamefont {D.}~\bibnamefont {Leuenberger}}, \bibinfo {author} {\bibfnamefont {A.}~\bibnamefont {Kemper}}, \bibinfo {author} {\bibfnamefont {J.}~\bibnamefont {Analytis}}, \bibinfo {author} {\bibfnamefont {I.}~\bibnamefont {Fisher}}, \bibinfo {author} {\bibfnamefont {P.}~\bibnamefont {Kirchmann}}, \bibinfo {author} {\bibfnamefont {T.}~\bibnamefont {Devereaux}}, \ and\ \bibinfo {author} {\bibfnamefont {Z.-X.}\ \bibnamefont {Shen}},\ }\href {\doibase 10.1016/j.elspec.2014.01.005} {\bibfield  {journal} {\bibinfo  {journal} {Journal of Electron Spectroscopy and Related Phenomena}\ }\textbf {\bibinfo {volume} {195}},\ \bibinfo {pages} {249} (\bibinfo {year} {2014}{\natexlab{b}})}\BibitemShut {NoStop}%
\bibitem [{\citenamefont {Na}\ \emph {et~al.}(2020)\citenamefont {Na}, \citenamefont {Boschini}, \citenamefont {Mills}, \citenamefont {Michiardi}, \citenamefont {Day}, \citenamefont {Zwartsenberg}, \citenamefont {Levy}, \citenamefont {Zhdanovich}, \citenamefont {Kemper}, \citenamefont {Jones} \emph {et~al.}}]{na2020establishing}%
  \BibitemOpen
  \bibfield  {author} {\bibinfo {author} {\bibfnamefont {M.}~\bibnamefont {Na}}, \bibinfo {author} {\bibfnamefont {F.}~\bibnamefont {Boschini}}, \bibinfo {author} {\bibfnamefont {A.~K.}\ \bibnamefont {Mills}}, \bibinfo {author} {\bibfnamefont {M.}~\bibnamefont {Michiardi}}, \bibinfo {author} {\bibfnamefont {R.~P.}\ \bibnamefont {Day}}, \bibinfo {author} {\bibfnamefont {B.}~\bibnamefont {Zwartsenberg}}, \bibinfo {author} {\bibfnamefont {G.}~\bibnamefont {Levy}}, \bibinfo {author} {\bibfnamefont {S.}~\bibnamefont {Zhdanovich}}, \bibinfo {author} {\bibfnamefont {A.~F.}\ \bibnamefont {Kemper}}, \bibinfo {author} {\bibfnamefont {D.~J.}\ \bibnamefont {Jones}},  \emph {et~al.},\ }\href@noop {} {\bibfield  {journal} {\bibinfo  {journal} {Physical Review B}\ }\textbf {\bibinfo {volume} {102}},\ \bibinfo {pages} {184307} (\bibinfo {year} {2020})}\BibitemShut {NoStop}%
\bibitem [{\citenamefont {Longa}\ \emph {et~al.}(2024)\citenamefont {Longa}, \citenamefont {Parent}, \citenamefont {Frimpong}, \citenamefont {Armanno}, \citenamefont {Gauthier}, \citenamefont {Légaré}, \citenamefont {Boschini},\ and\ \citenamefont {Jargot}}]{Longa2024}%
  \BibitemOpen
  \bibfield  {author} {\bibinfo {author} {\bibfnamefont {A.}~\bibnamefont {Longa}}, \bibinfo {author} {\bibfnamefont {J.-M.}\ \bibnamefont {Parent}}, \bibinfo {author} {\bibfnamefont {B.~K.}\ \bibnamefont {Frimpong}}, \bibinfo {author} {\bibfnamefont {D.}~\bibnamefont {Armanno}}, \bibinfo {author} {\bibfnamefont {N.}~\bibnamefont {Gauthier}}, \bibinfo {author} {\bibfnamefont {F.}~\bibnamefont {Légaré}}, \bibinfo {author} {\bibfnamefont {F.}~\bibnamefont {Boschini}}, \ and\ \bibinfo {author} {\bibfnamefont {G.}~\bibnamefont {Jargot}},\ }\href {\doibase 10.1364/OE.525265} {\bibfield  {journal} {\bibinfo  {journal} {Optics Express}\ }\textbf {\bibinfo {volume} {32}},\ \bibinfo {pages} {29549} (\bibinfo {year} {2024})}\BibitemShut {NoStop}%
\bibitem [{\citenamefont {Gauthier}\ \emph {et~al.}(2021)\citenamefont {Gauthier}, \citenamefont {Sobota}, \citenamefont {Pfau}, \citenamefont {Gauthier}, \citenamefont {Soifer}, \citenamefont {Bachmann}, \citenamefont {Fisher}, \citenamefont {Shen},\ and\ \citenamefont {Kirchmann}}]{Gauthier2021}%
  \BibitemOpen
  \bibfield  {author} {\bibinfo {author} {\bibfnamefont {N.}~\bibnamefont {Gauthier}}, \bibinfo {author} {\bibfnamefont {J.~A.}\ \bibnamefont {Sobota}}, \bibinfo {author} {\bibfnamefont {H.}~\bibnamefont {Pfau}}, \bibinfo {author} {\bibfnamefont {A.}~\bibnamefont {Gauthier}}, \bibinfo {author} {\bibfnamefont {H.}~\bibnamefont {Soifer}}, \bibinfo {author} {\bibfnamefont {M.~D.}\ \bibnamefont {Bachmann}}, \bibinfo {author} {\bibfnamefont {I.~R.}\ \bibnamefont {Fisher}}, \bibinfo {author} {\bibfnamefont {Z.~X.}\ \bibnamefont {Shen}}, \ and\ \bibinfo {author} {\bibfnamefont {P.~S.}\ \bibnamefont {Kirchmann}},\ }\href {\doibase 10.1063/5.0053479} {\bibfield  {journal} {\bibinfo  {journal} {Review of Scientific Instruments}\ }\textbf {\bibinfo {volume} {92}},\ \bibinfo {pages} {123907} (\bibinfo {year} {2021})}\BibitemShut {NoStop}%
\bibitem [{\citenamefont {Li}\ and\ \citenamefont {Persson}(2015)}]{li2015thermal}%
  \BibitemOpen
  \bibfield  {author} {\bibinfo {author} {\bibfnamefont {S.}~\bibnamefont {Li}}\ and\ \bibinfo {author} {\bibfnamefont {C.}~\bibnamefont {Persson}},\ }\href@noop {} {\bibfield  {journal} {\bibinfo  {journal} {Journal of Applied Mathematics and Physics}\ }\textbf {\bibinfo {volume} {3}},\ \bibinfo {pages} {1563} (\bibinfo {year} {2015})}\BibitemShut {NoStop}%
\bibitem [{\citenamefont {Zheng}\ \emph {et~al.}(2022)\citenamefont {Zheng}, \citenamefont {Zhao}, \citenamefont {Li}, \citenamefont {Chen}, \citenamefont {Zhu}, \citenamefont {Liu}, \citenamefont {Tang}, \citenamefont {Wang}, \citenamefont {Wang},\ and\ \citenamefont {Li}}]{Zheng2022}%
  \BibitemOpen
  \bibfield  {author} {\bibinfo {author} {\bibfnamefont {Y.}~\bibnamefont {Zheng}}, \bibinfo {author} {\bibfnamefont {H.}~\bibnamefont {Zhao}}, \bibinfo {author} {\bibfnamefont {W.}~\bibnamefont {Li}}, \bibinfo {author} {\bibfnamefont {Z.}~\bibnamefont {Chen}}, \bibinfo {author} {\bibfnamefont {W.}~\bibnamefont {Zhu}}, \bibinfo {author} {\bibfnamefont {X.}~\bibnamefont {Liu}}, \bibinfo {author} {\bibfnamefont {Q.}~\bibnamefont {Tang}}, \bibinfo {author} {\bibfnamefont {H.}~\bibnamefont {Wang}}, \bibinfo {author} {\bibfnamefont {C.}~\bibnamefont {Wang}}, \ and\ \bibinfo {author} {\bibfnamefont {Z.}~\bibnamefont {Li}},\ }\href {\doibase 10.1021/acs.nanolett.2c01105} {\bibfield  {journal} {\bibinfo  {journal} {Nano Lett.}\ }\textbf {\bibinfo {volume} {22}},\ \bibinfo {pages} {6102} (\bibinfo {year} {2022})}\BibitemShut {NoStop}%
\end{thebibliography}
\end{document}